\documentclass[12pt, a4paper]{article}

\usepackage{geometry}
\usepackage{lineno}
\usepackage{amsmath, amssymb, amsfonts, appendix}
\usepackage{amsthm}
\usepackage{extarrows}

\usepackage{enumerate}

\usepackage{graphicx}
\usepackage{eepic,color,colordvi,amscd}
\usepackage{subcaption}

\newtheorem{theorem}{Theorem}[section]
\newtheorem{lemma}[theorem]{Lemma}
\newtheorem{definition}[theorem]{Definition}

\newtheorem{problem}[theorem]{Problem}

\newtheorem{remark}[theorem]{Remark}
\newtheorem{corollary}[theorem]{Corollary}
\newtheorem{proposition}[theorem]{Proposition}

\newcommand{\R}{\mathbb{R}}
\newcommand{\Z}{\mathbb{Z}}

\newcommand{\im}{\operatorname{Im}}
\newcommand{\inn}[1]{\langle #1\rangle}
\newcommand{\rs}{\setminus}
\newcommand{\supp}{\operatorname{supp}}
\newcommand{\mc}[1]{\mathcal{#1}}

\newcommand{\Ftwo}{\mathbb F_2}
\newcommand{\wt}{\operatorname{wt}}
\newcommand{\one}{\mathbf 1}
\newcommand{\cZ}{\mathcal Z}
\newcommand{\cB}{\mathcal B}

\newcommand{\exc}{\operatorname{exc}_2}
\newcommand{\rex}{\operatorname{rex}_2}
\newcommand{\rcex}{\operatorname{rcex}}
\newcommand{\rctr}{\operatorname{rctr}}
\newcommand{\cycexc}{\operatorname{cex}}
\newcommand{\ctr}{\operatorname{ctr}}
\newcommand{\GL}{\operatorname{GL}}
\newcommand{\MaxCut}{\operatorname{MaxCut}}

\newcommand{\rank}{\operatorname{rank}}

\newcommand{\F}{\mathbb{F}}

\usepackage{graphicx}
\usepackage{tikz} 
\usepackage{hyperref}
\hypersetup{
	colorlinks=true,
	linkcolor=blue,
	filecolor=magenta,      
	urlcolor=cyan,
	pdftitle={},
}

\title{Binary Affine Optimization with Nearly-Graphic Constraint Matrices}
\author{
	Qiming Fang\thanks{Beijing International Center for Mathematical Research, Peking University, Beijing 100871, China. Email: \texttt{fangqiming@bicmr.pku.edu.cn}}
	\and
	Sihong Shao\thanks{CAPT, LMAM and School of Mathematical Sciences, Peking University, Beijing 100871, China. Email: \texttt{sihong@math.pku.edu.cn}}
	\and
	Yuxuan Wu\thanks{School of Mathematical Sciences, Peking University, Beijing 100871, China. Email: \texttt{2501110053@stu.pku.edu.cn}}
}
\date{}

\begin{document}
\maketitle
	{\small {\textbf{Abstract}:} 	We present a polynomial-time algorithm for binary affine optimization with a linear objective when every column of the constraint matrix has Hamming weight at most 2, and prove that the problem is NP-hard when every column of the constraint matrix has Hamming weight exactly 3. We obtain fixed-parameter tractable (FPT) algorithms for the unrestricted problem provided that a \emph{nearly-graphic certificate} is given, where the certificate roughly measures how far the constraint matrix is from having columns of Hamming weight at most 2. We apply our FPT algorithm to Max-Lin\(_2\) on graphs and obtain FPT algorithms parameterized by the \emph{contraction distance} and \emph{contraction rank} of the graph, utilizing the concept of \emph{Eulerian-spanning set}. In the case of contraction distance, another FPT algorithm can be used to find the corresponding certificate; that is, we obtain an FPT algorithm that requires no information beyond the problem instance itself.	
		
		
	}

	
	\section{Introduction}
	\label{s1}
	
	Many exact optimization, decoding, and graph-labeling problems over binary
	variables can be written as, or converted to the following NP-hard binary affine optimization, with a linear objective. (For the notations in the introduction, one can look up them in ``Basic assumptions" at the end of it.)
	\begin{problem}[binary affine optimization with a linear objective, or BAO]\label{prob:linear}
		Given $A\in\Ftwo^{r\times m},b\in\Ftwo^r,\Delta\in\R^m$, solve the following problem and find an optimal assignment:
		\begin{equation} \label{eq:rank-normalized}
			\min\left\{\Delta^{\mathsf T}x: Ax=b,\quad x\in\Ftwo^m\right\},
		\end{equation}
		where $\Delta^{\mathsf T}x$ is evaluated in $\R$, and elements $0,1\in\Ftwo$ are regarded as $0,1\in\R$, respectively.
	\end{problem}

	Special cases of BAO are closely connected with classical problems in coding theory and with optimization problems for linear systems over $\Ftwo$. If $Ax=b$ is feasible, then its solution set is an affine coset of the linear code $\ker A$. In the special case $\Delta_j=1$ for all $j$, BAO is exactly the problem of finding a minimum-Hamming-weight representative of a prescribed coset, i.e., the syndrome-decoding or coset-leader problem for a binary linear code. The general decoding problem was shown to be NP-complete \cite{BerlekampMcElieceVanTilborg1978}. Related minimization and approximability questions for linear systems, including the minimization of the number of nonzero variables and binary-restricted variants, were studied systematically \cite{AmaldiKann1998}.
	
%
%

Weighted systems of linear equations over \(\Ftwo\), usually formulated as Max-Lin\(_2\), can also be converted into the form of BAO and provide another closely related viewpoint. Max-Lin\(_2\) contains several graph optimization problems, such as MaxCut and signed frustration. Parameterized complexity results for Max-Lin\(_2\) and its sparse-occurrence cases were developed \cite{maxlin2,CrowstonEtAl2013}, where the approximation problem was  considered. The W[1]-hardness of Max-Lin\(_2\), parameterized by the total weight of violated equations, was proved \cite{maxlin2}. The problem of finding an assignment for which the total weight of satisfied equations exceeds that of random guessing was investigated \cite{CrowstonEtAl2013},
where the parameter is the gap from half the total weight and the special case in which every equation contains at most \(r\) variables was studied. 


	In this paper, we study the complexity of BAO, present fixed-parameter tractable (FPT) algorithms, and explore their applications to Max-Lin\(_2\) (where we allow real weights). As far as we are aware, both our approach and our results are structurally new. For the BAO problem, our results focus on the columns of the constraint matrix whose Hamming weight is at least three, which, to the best of our knowledge, have not been considered elsewhere. For Max-Lin\(_2\), we consider the exact problem rather than the approximation problem. We convert Max-Lin\(_2\) into the form of BAO and solve the latter, rather than attempting to solve it directly as in \cite{maxlin2,CrowstonEtAl2013}.

	
	
	We now describe the results, techniques, and key concepts of this work. We first show that if every column of \(A\) has Hamming weight at most \(2\), then BAO can be solved in polynomial time. For convenience, we say that \(A\)  or, equivalently, the BAO problem, satisfies the \emph{weight-two condition} when this holds.
	
		\begin{theorem}\label{thm:graphic-core}
		Let $A\in\Ftwo^{r\times m}$ be a matrix satisfying the weight-two condition, and let $b\in\Ftwo^r$, $w\in\R^m$.  Then the feasibility
		of $Ax=b$ can be determined in polynomial time.  If the system is feasible, both the value
		\begin{equation}
			\min\{w^{\mathsf T}x:Ax=b,\ x\in\Ftwo^m\}
		\end{equation}
		and an optimal vector $x$ can be computed in polynomial time.
	\end{theorem} \par
	A constraint matrix satisfying weight-two condition has a graph-incidence interpretation. A BAO instance with such a constraint matrix can be converted to a minimum-weight $T$-join problem on a graph, for which classical matching methods provide the algorithmic foundation
	\cite{EdmondsJohnson1973}. 
	
	On the other hand, we show that even when every column of $A$ has Hamming weight exactly 3, BAO is still NP-hard. This establishes the tightness of the weight-two condition. 
	\begin{theorem}
		\label{thm:weight-three}
		The following problem is NP-complete: given $A\in\Ftwo^{r\times m}$ in which
		every column has Hamming weight three, and an integer $q\in\mathbb Z_{\ge0}$, decide
		whether there is $x\in\Ftwo^m$ such that
		\[
		Ax=\one,\qquad \wt(x)\le q.
		\]
		Consequently, minimizing the Hamming weight $\wt(x)$ subject to $Ax=\one$ is NP-hard even when every column of $A$ has Hamming weight three.
	\end{theorem} \par
	The proof of Theorem \ref{thm:weight-three} is in Appendix \ref{sec:hard}. \par
	With Theorem \ref{thm:graphic-core}, we can develop FPT algorithms for general BAO. We start from a simple application. Let $J$ be the indexes of columns of $A$ with Hamming weight at least 3. We enumerate all possibilities of $x_J$. After $x_J$ is fixed, the remaining of the constraint is of the form $A_{\overline J}x_{\overline J}=b'$, where $A_{\overline J}$ satisfies the weight-two condition. Therefore the remaining problem can be solved within polynomial time, and the total running time must be $O(2^{|J|}\max\{r,m\}^{O(1)})$. However, we can apply this algorithm with some twist. First, for any $U\in\GL(r,2)$, $Ax=b\Leftrightarrow UAx=Ub$. Therefore we can pick any $U\in\GL(r,2)$ and instead consider the columns of $UA$ of Hamming weight at least 3. By applying the row operation $U$ to $A$, the number of columns of Hamming weight at least 3 can be reduced. This leads to the concept of \emph{nearly-graphic certificate}.  
	\begin{definition}
		Let $A\in\Ftwo^{r\times m}$. A pair $(U,J), U\in\GL(r,2),J\subseteq [m]$ satisfying $\forall j\in [m]\backslash J, \wt((UA)_{j})\le2$ is called a nearly-graphic certificate for $A$.  
	\end{definition} \par
	Let $\mathcal{C}_A$ collect all nearly-graphic certificates for $A$. From now on, if $A$ is clear from the context, we will simply call $(U,J)$ a nearly-graphic certificate. The word \emph{nearly-graphic} comes from the fact that a binary matrix satisfying the weight-two condition can be converted to the incidence matrix of some graph. For the enumeration of coordinates in $J$, we may not have to go through all $2^{|J|}$ possibilities of $x_J$. Instead, we can consider $\im ((UA)_J)$, enumerate all elements in $\im ((UA)_J)$, and compute an optimal $x_J$ for each of the images. We have $|\im ((UA)_J)|=2^{\rank((UA)_J)}$. Moreover, $\rank((UA)_J)=\rank(A_J)\leq|J|$, which is an advantage. This leads to the definition of the following \emph{nearly-graphic parameters}.
	\begin{definition}[nearly-graphic parameters]
		Let $A\in\Ftwo^{r\times m}$. Define the following parameters:
		\begin{align}
			\exc(A)
			&=\min_{(U,J)\in\mathcal{C}_A}|J|,
			\label{eq:exception-number}\\
			\rex(A)
			&=\min_{(U,J)\in\mathcal{C}_A}\rank(A_J).
		\end{align}
		If $(U,J)$ is a nearly-graphic certificate for $A$, then its support size is $|J|$, and its non-graphic rank is
		\begin{equation}
			\rho(U,J)=\rank((UA)_J)=\rank(A_J).
			\label{eq:certificate-rank}
		\end{equation} \par
	\end{definition} \par
	The notations exc, rex are abbreviations of the phrases ``exceptional columns" and ``rank of exceptional columns", respectively. When a nearly-graphic certificate $(U,J)$ for $A$ is provided, we obtain two FPT algorithms solving BAO, parameterized by $|J|$ and $\rank(A_J)$, respectively.
	\begin{theorem}\label{thm:affine}
		Let $(A,b,\Delta)$ be an instance of BAO, and $N=\max\{r,m\}$. Suppose a nearly-graphic certificate $(U,J)$ is given.  Put
		$k=|J|$ and $\rho=\rank(A_J)=\rank((UA)_J)$. The validity of the supplied certificate can be checked in time polynomial in $N$. If
		it is valid, then there are two algorithms finding the optimum and an optimal assignment in $O(2^\rho N^{O(1)})$ and $O(2^k N^{O(1)})$ time, respectively.
	\end{theorem} \par
	We explore the application of Theorem \ref{thm:affine} and consider the weighted Max-Lin$_2$ problem, which is defined as follows.
	\begin{definition}[weighted Max-Lin$_2$]
		Let $M\in\Ftwo^{q\times n}, d\in\Ftwo^q, w\in\R^q$. We aim to compute
		$$\min\{w^{\mathsf T}z:z=Mx+d,\ z\in\Ftwo^q\},$$
		and an optimal vector $z$. Here $w^{\mathsf T}z$ is evaluated in $\R$, and elements $0,1\in\Ftwo$ are regarded as $0,1\in\R$, respectively.
	\end{definition} \par
	The form of Max-Lin$_2$ is different from that of BAO, hence we need to convert it. This is done by computing a representation of the left-nullspace of $M$. Specifically, let $H$ be a matrix whose row space is $\ker M^{\mathsf T}$ (by $\ker$ we always mean the right-nullspace of a matrix, that's why the left-nullspace of $M$ reads $\ker M^{\mathsf T}$ here). Then we have 
	\begin{equation}
		\{Mx+d:x\in\Ftwo^n\}
		=\{z\in\Ftwo^q:Hz=Hd\}.
		\label{eq:maxlin-coset}
	\end{equation}
	Indeed, if $z=Mx+d$, then $HM=0$, and hence $Hz=Hd$.  Conversely, if $Hz=Hd$, then $z+d\in\ker H$.  Since the row space of $H$ is $\ker M^{\mathsf T}$,
	\[
	\ker H=(\ker M^{\mathsf T})^\perp=\im M.
	\]
	Thus $z+d=Mx$ for some $x$, which proves \eqref{eq:maxlin-coset}. Weighted Max-Lin$_2$ is therefore equivalent to
	\begin{equation}
		\min\{w^{\mathsf T}z:Hz=Hd,\ z\in\Ftwo^q\}.
		\label{eq:maxlin-decoding}
	\end{equation}
	By the conversion, we can recover an optimal assignment of $x$ if we know the optimal assignment of \eqref{eq:maxlin-decoding} in polynomial time. Therefore, if a nearly-graphic certificate $(U,J)$ of $H$ is provided, we can apply Theorem \ref{thm:affine} and solve Max-Lin$_2$ in $2^{\rank(H_J)} N^{O(1)}$, or $2^{|J|} N^{O(1)}$ time. \par
	However, the choice of $H$ is not unique in general. Therefore, if a nearly-graphic certificate is to be provided, one must first specify $H$, which is unrealistic. Hence we make further exploration by considering a special case of Max-Lin$_2$, i.e. Max-Lin$_2$ on graphs. For brevity, here we omit the original form of Max-Lin$_2$ on graphs, which is in Section \ref{sec:maxlingraph}. We will show that, for Max-Lin$_2$ on a graph $G=(V,E)$, $H$ must be taken to represent the cycle space of $G$. This leads to the concept of \emph{Eulerian-spanning set} of $G$, which is a multiset of vectors spanning the cycle space of $G$. Therefore we can take $H$ to be the incidence matrix of a Eulerian-spanning set $\cB$ of $G$. Denote the matrix by $A_\cB$ and we get the following BAO problem:
	$$\min\{w^{\mathsf T}z:A_{\cB}z=A_{\cB}c,\ z\in\Ftwo^E\}.$$
	We will show that, $\exc(A_\cB)$, $\rex(A_\cB)$ are independent from the choice of $\cB$, and are equal to the \emph{contraction distance} and \emph{contraction rank} of $G$, respectively. The contraction distance is the minimum number of edges to contract to make $G$ planar, while the contraction rank is a little more complicated and will be explained in Section \ref{sec:parameters}. Moreover, $J$ can be simply taken to be the set of edges whose contraction makes $G$ planar. More importantly, we can use the FPT algorithms in \cite{GolovachEtAl2013,AbelloEtAl2017} to compute the contraction distance of $G$, and find the corresponding edge set. Therefore we get an FPT algorithm for Max-Lin$_2$ on graphs parameterized by contraction distance, which needs no information other than the problem itself.
	\begin{theorem}
		\label{thm:intrinsic-contraction-fpt}
		Let $G=(V,E)$ be a graph, and $\ctr(G)$ be the contraction distance of $G$. One can solve weighted Boolean Max-Lin$_2$ on a graph $G=(V,E)$, and find an optimal assignment in time
		$F(\ctr(G))\max\{|V|,|E|\}^{O(1)}$, where $F$ is a function independent of $|V|,|E|$.
	\end{theorem} 
	

	A formal definition of $\ctr(G)$ can also be found in Section~\ref{sec:parameters}. We explore applications of our algorithms for Max-Lin\(_2\) to specific graph optimization problems. Specifically, we consider MaxCut and signed frustration, to which Theorem~\ref{thm:intrinsic-contraction-fpt} can be applied directly. Moreover, when the edge set $J$ is provided, we design a more refined algorithm that reduces the exponent with respect to $\max\{|V|,|E|\}$. This algorithm generalizes the result of \cite{my}, which considers nonnegative weights, to the case of arbitrary real weights. A further application of this algorithm yields an FPT algorithm for Max-Lin\(_2\) parameterized by the crossing number, provided that a corresponding drawing is given.

	As mentioned earlier, FPT algorithms exist for computing the contraction distance of $G$. However, we are not aware of any FPT algorithm for computing the contraction rank of $G$, even though the latter is generally smaller and thus has the potential to yield faster algorithms. We investigate the possible separation between the contraction distance and the contraction rank of $G$. In Appendix~\ref{separation}, we show that the contraction distance can be arbitrarily larger than the contraction rank.

	
	\paragraph{Paper organization} 
	Section~\ref{sec:tjoin} gives the polynomial algorithms for BAO under the weight-two condition. Section~\ref{sec:FPT} gives the two FPT algorithms for BAO. Section~\ref{sec:maxlingraph} converts Max-Lin$_2$ on graphs to BAO using the concept of Eulerian-spanning set. Section~\ref{sec:parameters} characterizes the nearly-graphic parameters of $A_\cB$ by the contraction rank and the contraction distance, and gives FPT algorithms for solving Max-Lin$_2$ on graphs.

	\paragraph{Basic assumptions} 
	\begin{itemize}
		\item Every graph is assumed to be finite and undirected. Graphs are allowed to have multiple edges and loops. By simple graph we mean graphs without multiple edges and loops. If $G$ is a graph, then $c(G)$ denotes the number of connected components of $G$. If $G=(V,E)$ is a graph and $V'\subseteq V$, then $G[V']$ is the subgraph induced by $V'$.
		\item Copies of the same element in a multiset are treated differently. For an integer $t\ge0$, write
		$[t]=\{1,\ldots,t\}$, with the convention $[0]=\varnothing$.
		\item Unless stated otherwise, all vector spaces, matrices, and linear-algebraic operations in Sections \ref{sec:tjoin}--\ref{sec:parameters} are
		over $\Ftwo$. For a binary vector $x$, let
		$\supp(x)$ and $\wt(x)$ denote its support and Hamming weight.  For $J\subseteq[m]$ where $m$ is clear from the context, write $\overline J=[m]\setminus J$ and use $x_J$
		and $A_J$ for the corresponding subvectors and column submatrices. The notation
		$A_{j}$ denotes the $j$th column of $A$, and $\one$ denotes the all-ones vector
		of the dimension determined by context. The rank of an empty matrix is zero. 
		\item For the vector space $\Ftwo^n,n\in\Z^+$, the non-degenerate symmetric bilinear form $\inn{\cdot,\cdot}$ is taken in the obvious way, i.e. $\inn{x,y}=\sum_{i\in[n]}x_iy_i$. Orthogonal subspaces are taken with respect to this bilinear form.
	\end{itemize}
	
	\section{Polynomial algorithm under weight-two condition}
	\label{sec:tjoin}
	
	To solve BAO under weight-two condition, we first convert the problem to a T-join problem, and then solve the latter in polynomial time. 
	
	\begin{definition}
		Let $G=(V,E)$ be a graph. For $Q\subseteq E$, let $\partial_G Q$ be the set of vertices of odd degree in the subgraph $(V,Q)$. A loop contributes two to the degree. For $T\subseteq V$, a set $Q\subseteq E$ is a $T$-join if $\partial_G(Q)=T$.
	\end{definition}

	Since a loop contributes $2$ to the degree, including it or not does not affect $\partial_G(Q)$. Hence, we may select all negative-weight loops, delete all loops from the graph, and work with the resulting graph instead. Therefore, when considering $T$-joins, we may assume without loss of generality that $G$ is loopless. It is easy to see that a $T$-join exists if and only if every connected component of $G$ contains an even number of vertices of $T$. Necessity follows from a simple counting argument. For sufficiency, pair the vertices of $T$ arbitrarily within each component, choose a path for each pair, and take the symmetric difference of the edge sets of these paths.

	
	The $T$-join problem can be solved in polynomial time. Moreover, if $G$ is planar, the running time can be further decreased. The algorithm for planar graphs will be used later. 
	
	\begin{lemma} \label{lem3main}
		Let \(G=(V,E)\) be a loopless graph with weight function \(f\colon E\to\mathbb{R}\), let \(V_t\subseteq V\), and write \(n=|V|\), \(m=|E|\). Then there is an \(O(\max\{m,n\}^3)\)-time algorithm that either finds a minimum-weight edge subset \(E'\subseteq E\) satisfying \(\partial_G(E')=V_t\), or reports that none exists. If \(G\) is planar, the running time improves to \(O(\max\{m,n\}^{3/2}\log \max\{m,n\})\).
		%
	\end{lemma}
	
	The proof of Lemma~\ref{lem3main} is delayed to Appendix \ref{aa}.  Next we will present the proof of Theorem~\ref{thm:graphic-core} which shows how to convert Problem \ref{prob:linear} to a $T$-join problem and then solve it under the weight-two condition.
	
	
	\begin{proof}[Proof of Theorem~\ref{thm:graphic-core}]
		Because zero columns do not contribute to the affine system, we can simply pick the value of $x$ on those coordinates to be 0 if the corresponding weight is positive, and 1 otherwise. From now on we assume that $A$ has no zero columns. Construct a graph $G_A$ with vertex set $[r]\cup\{\star\}$, where $\star$ is a new vertex.  A column with support $\{u, v\}$ becomes an edge $uv$, and a column with support $\{u\}$ becomes an edge $u\star$.  The edge inherits the weight of the corresponding column.
		
		Set
		\begin{equation}
			T_b=\supp(b)\mathbin{\triangle}
			\begin{cases}
				\{\star\},&\text{if }\wt(b)\text{ is odd},\\
				\varnothing,&\text{if }\wt(b)\text{ is even}.
			\end{cases}
			\label{eq:target-T}
		\end{equation}
		For $x\in\Ftwo^m$, let $E_x$ be the corresponding set of graph edges. Then we have $\forall x, Ax=(\partial_{G_A}E_x)_{[r]}$.  Hence
		\[
		Ax=b \quad\Longleftrightarrow\quad \partial_{G_A}E_x=T_b.
		\]
		By Lemma \ref{lem3main}, we can solve the above problem and recover an optimal vector $x$ in polynomial time.
	\end{proof}
	\begin{remark}
		What we do in the proof is equivalent to attaching a row to $A$ that is the sum of all rows in $A$. If we denote the matrix obtained by $A'$, then every column of $A'$ has Hamming weight two. Moreover, $A'$ is exactly the incidence matrix of the graph $G_A$.
	\end{remark}
	
	\section{FPT algorithm with nearly-graphic parameters}\label{sec:FPT}
	We first need a lemma that solves the subproblem of specific coordinates.
	
	\begin{lemma}[subproblem of specific coordinates]
		\label{lem:syndrome-dp}
		Let $C\in\Ftwo^{r\times m}, J\subseteq[m], \Delta\in\R^m$.  Put $k=|J|$ and
		$\rho=\rank(C_J)$.  For every $s\in\im(C_J)$, define
		\begin{equation}
			D_J(s)=\min\{\Delta_J^{\mathsf T}\alpha:
			C_J\alpha=s,\ \alpha\in\Ftwo^J\}.
			\label{eq:syndrome-cost}
		\end{equation}
		Let $N=\max\{r,m\}$. Then all values $D_J(s)$ for $s\in\im(C_J)$, together with all minimizers $\alpha$, can be computed in time $O(2^\rho N^{O(1)})$.
	\end{lemma}
	
	\begin{proof}[Proof of Lemma~\ref{lem:syndrome-dp}]
		Order $J$ as $j_1,\ldots,j_k$.  Choose a linear isomorphism
		$f:\im(C_J)\mapsto\Ftwo^\rho$, which can be computed by the Gaussian elimination, and
		let $z_i=f(C_{j_i})$. For
		$i\in\{0,\ldots,k\}$ and $t\in\Ftwo^\rho$, let
		$$D_i(t)=\min\{\Delta_J^{\mathsf T}\alpha:C_J\alpha=f^{-1}(t),\supp(\alpha)\subseteq \{j_1,\ldots,j_i\}\}.$$
		Then $D_0(0)=0$ and $D_0(t)=+\infty$ for $t\ne0$.  Splitting a subset
		according to whether it contains $j_i$ gives
		\begin{equation}
			D_i(t)=\min\bigl\{D_{i-1}(t),
			D_{i-1}(t+z_i)+\Delta_{j_i}\bigr\}.
			\label{eq:syndrome-dp}
		\end{equation}
		We have $\forall t\in\Ftwo^\rho, D_k(t)=D_J(f^{-1}(t))$. Then one can compute all values of $D_J(f^{-1}(t))$, and hence $D_J(s)$ in time $O(2^\rho N^{O(1)})$. Moreover, we can keep track of the optimal assignments achieving $D_i(t)$ during the computation. Therefore those $\alpha$ achieving $D_J(s)$ can be simultaneously obtained in the end of the induction.
	\end{proof} 
	
	\begin{proof}[Proof of Theorem~\ref{thm:affine}]
		First apply the Gaussian Elimination to the system $Ax=b$. If it detects an inconsistent system, report inconsistency and stop. Then we can verify in polynomial time that $U$ is invertible and that every column of $UA$ outside
		$J$ has weight at most two. Afterwards, set
		$C=UA$ and $g=Ub$, so that $Cx=g$ is equivalent to $Ax=b$. Now we give two different algorithms that achieve the two time complexities respectively.\par
		(1) For time complexity $O(2^\rho N^{O(1)})$, we use the following algorithm to solve the problem. Use Lemma \ref{lem:syndrome-dp} to compute all $D_J(s)$, together with those $x_J$ achieving the optimum. The time cost of this procedure is $O(2^\rho N^{O(1)})$. If $C_Jx_J=s$, then the remaining coordinates must satisfy
		\begin{equation}
			C_{\overline J}x_{\overline J}=g+s.
			\label{eq:syndrome-branch}
		\end{equation}
		We note that addition and subtraction are equivalent in $\Ftwo$, therefore $g+s=g-s$. Moreover, $C_{\overline J}$ has column weight at most two, so we can apply
		Theorem \ref{thm:graphic-core} to either detect infeasibility or obtain an $x_{\overline J}$ such that $\Delta_{\overline J}^{\mathsf T} x_{\overline J}$ achieves minimum. Denote this minimum by $Q(s)$, and take $Q(s)=+\infty$ when the system is infeasible.  Then it is easy to see that the optimum value for the original BAO problem is
		\begin{equation}
			\min_{s\in\im(C_J)}\bigl(D_J(s)+Q(s)\bigr).
			\label{eq:rank-decomposition}
		\end{equation} 
		If $Q(s)=+\infty$ for every $s\in\im(C_J)$, report that the affine system is
		infeasible.  Otherwise, choose an $s$ attaining a finite minimum in
		\eqref{eq:rank-decomposition}. We already have $x_J$ after applying the algorithm in Lemma \ref{lem:syndrome-dp}. $x_{\overline J}$ can be recovered when applying Theorem \ref{thm:graphic-core}. Concatenate $x_J$ and $x_{\overline J}$ to obtain an optimal $x$. \par
		(2) For time complexity $O(2^k N^{O(1)})$, we use the following algorithm. For every $x_J$, use Theorem \ref{thm:graphic-core} to compute 
		$$Q(x_J)=\min\{\Delta_{\overline J}^{\mathsf T} x_{\overline J}: C_{\overline J}x_{\overline J}=g+C_Jx_J\}.$$
		Then the optimal value is equal to 
		$$\min_{x_J}\{\Delta_J^{\mathsf T}x_J+Q(x_J)\}$$
		and the total time complexity is $O(2^k N^{O(1)})$.
	\end{proof}
	
	We would like to point out that Theorem~\ref{thm:affine} is certificate-based.  It verifies a supplied certificate and optimizes from it, but it does not find a certificate attaining $\rex(A)$.
	
	\section{\texorpdfstring{Max-Lin$_2$}{Max-Lin2} on graphs and Eulerian-spanning set}\label{sec:maxlingraph}
	
	Before we proceed, we make the following conventions. Suppose $S$ is a finite multiset and $\F$ is a field. Then $\F^{S}$ is the linear space over $\F$ with dimension $|S|$. $\F^{S}$ is regarded to be indexed by the multiset $S$. Moreover, if $\F=\Ftwo$, then every element in $\Ftwo^S$ is regarded as an indicator function of some subset of $S$. Suppose $S_1,S_2$ are two subsets of the same multiset $S$. Then $S_1\triangle S_2$ is the symmetric difference between $S_1,S_2$. Moreover, the symmetric bilinear form $\inn{\cdot, \cdot}$ on $\Ftwo^S$ is defined as $\inn{x, y}=\sum_{s\in S}x(s)y(s), \forall x, y\in \Ftwo^S$.
	It is obvious that $\inn{\cdot,\cdot}$ is non-degenerative.
	
	\begin{problem}[weighted Boolean Max-Lin$_2$ on graphs] \label{prob:weighted}
		Suppose $G=(V,E)$ is a graph in which each edge \(e\in E\) has weight \(w_e\in\mathbb{R}\). In addition, a violation variable $c_e\in\Ftwo$ is associated with each edge $e\in E$. For \(x\in\mathbb{F}_2^V\), the violation vector $z(x)\in\Ftwo^E$ is defined by
		\[
		\forall e\in E,\ e= (u,v),\ z(x)_{e}=x_u+x_v+c_e.
		\]
		Our goal is to solve the following optimization problem, and compute an optimal assignment:
		$$\min\{w^{\mathsf T}z(x):x\in\Ftwo^V\}.$$
	\end{problem} 
	
	\begin{remark}\label{rmk:loopless}
		A loop corresponds to a fixed bit \(c_e\). By adding the constant \(w_e c_e\) to the objective and deleting the loop equations, we may assume that \(G=(V,E)\) is loopless. Moreover, one can merge multiple edges, and modify the weights and violation variables accordingly to obtain an equivalent problem (up to an additive constant in the objective). Therefore we may assume that \(G=(V,E)\) is simple. The loopless assumption is used in Appendix \ref{ab}, in the proof of Theorem \ref{thm:faster}. The simple assumption is used in Appendix \ref{ab}, in application to the case of crossing numbers.
	\end{remark} 
	
	
	The violation vector \(z\) admits a more concise expression in terms of the incidence matrix of a graph. Given a graph \(G=(V,E)\), its incidence matrix \(M_G\in\Ftwo^{V\times E}\) is defined as follows: its rows are indexed by \(V\), and its columns are indexed by \(E\); moreover, for all \(v\in V\) and \(e\in E\), \(M_G(v,e)=1\) if and only if \(e\) is not a loop and \(v\) is an endpoint of \(e\).


	\begin{problem}[weighted Boolean Max-Lin$_2$ on graphs] \label{prob:concise}
		Suppose \(G=(V,E)\) is a graph, \(w\in\mathbb{R}^E\) is a weight vector, and \(c\in\mathbb{F}_2^E\) is a given violation vector. For \(x\in\mathbb{F}_2^V\), the violation vector \(z(x)\in\mathbb{F}_2^E\) is defined as
		\[
		z(x)=M_G^{\mathsf T}x+c.
		\]
		Our goal is to solve the optimization problem:
		\(
		\min\{w^{\mathsf T}z(x):x\in\mathbb{F}_2^V\}.
		\)
	\end{problem} \par
	It is easy to see that Problem \ref{prob:weighted} is equivalent to Problem \ref{prob:concise}. We will use the latter form from now on. Next we give the definition of Eulerian-spanning sets, and convert Max-Lin$_2$ to BAO.
	\begin{definition}
		Suppose $G=(V,E)$ is a graph and $M_G$ is its incidence matrix. Define the cycle space $\cZ(G)$ of $G$ to be $\cZ(G)=\ker M_G$. A Eulerian-spanning set of $G$ is a multiset
		$\cB=(B_1,\dots,B_r)$ consisting of elements in $\cZ(G)$, such that the span of elements in $\cB$ is $\cZ(G)$. If elements in $\cB$ are linearly independent, then $\cB$ is called a Eulerian basis of $G$. Its edge frequency for a certain $e\in E$ is
		\(
		f_{\cB}(e)=|\{i:e\in B_i\}|.
		\)
		Let $A_{\cB}\in\Ftwo^{r\times E}$ be the matrix whose $i$th row is the
		incidence vector of $B_i$.  Define the following exceptional parameters:
		\begin{equation}
			J_{\cB}=\{e\in E:f_{\cB}(e)\ge3\},
			\qquad
			\rho_{\cB}=\rank((A_{\cB})_{J_{\cB}}),
			\label{eq:basis-exception-rank}
		\end{equation}
		and $\rho_\cB$ will be called the exceptional rank of $\cB$.
	\end{definition}
	
	\begin{remark} \label{cyclebasis}
		Every Eulerian subgraph of $G$ is an edge-disjoint union of cycles: a nonempty Eulerian component contains
		a cycle, and deleting the edges of that cycle preserves even degree, so induction
		applies. Therefore, given any Eulerian basis $\mc{B}$, we can split every element into cycles, and then select a
		basis from the whole set of cycles. It is easy to see the above procedure does not increase any edge frequency.
	\end{remark} 
	
	Using the Eulerian-spanning set of $G$ and \eqref{eq:maxlin-decoding} yields 
	\begin{equation}
		\min\{w^{\mathsf T}z(x):x\in\Ftwo^V\}=\min\{w^{\mathsf T}z:A_{\cB}z=A_{\cB}c,\ z\in\Ftwo^E\}.
		\label{eq:graph-min}
	\end{equation}
	If we have the solution to the latter, then we are able to recover the solution to the former in polynomial time.
	In particular, the maxcut and signed frustration problems are special cases of Max-Lin$_2$ on graphs.
	
	\begin{problem}[maxcut]
		Let $G=(V,E)$ be a graph with edge-weight vector
		$w\in\R^E$, and \(W=\sum_{e\in E}w_e\). 
		Define \(\MaxCut_w(G)=\max_{x\in\Ftwo^V}\sum_{e=(u,v)\in E}w_e\,\mathbf 1_{\{x_u\ne x_v\}}\).
		Then 
		\[
		\MaxCut_w(G)=W-\min\{w^{\mathsf T}(M_G^{\mathsf T} x+\one):x\in\Ftwo^V\}.
		\] 		
	\end{problem}
	
	That is, the maxcut problem  corresponds to the case of $c=\one$ in Problem \ref{prob:concise}, and 
	applying \eqref{eq:graph-min} then yields the following equivalent form.

	\begin{corollary}\label{cor:maxcut}
		Let $G=(V,E)$ be a graph with edge-weight vector
		$w\in\R^E$, \(W=\sum_{e\in E}w_e\),  and let $\cB$ be a Eulerian-spanning set of $G$. Then
		\begin{equation}
			\MaxCut_w(G)
			=W-\min\{w^{\mathsf T}z:A_{\cB}z=A_{\cB}\one,\ z\in\Ftwo^E\}.
			\label{eq:maxcut}
		\end{equation}
	\end{corollary}
	
	
	The signed frustration problem involves the concept of a signed graph. A signed graph is a graph with each edge marked with a sign, positive or negative. For a signed graph, encode a positive edge by $c_e=0$ and a negative edge by $c_e=1$.  A signed graph is balanced exactly when every cycle contains an even number of negative edges \cite{Harary1953}.
	
	\begin{problem}[signed frustration]
		\label{cor:frustration}
		Let $G=(V,E)$ be a signed graph with sign vector
		$c\in\Ftwo^E$ (obtained in the way described above) and edge-weight vector $w\in\R^E$, and let $\cB$
		be a Eulerian basis of $G$.  Then the minimum total weight of violated edges in any vertex assignment is
		\[
		\min\{w^{\mathsf T}(M_G^{\mathsf T} x+c):x\in\Ftwo^V\}=\min\{w^{\mathsf T}z:A_{\cB}z=A_{\cB}c,\ z\in\Ftwo^E\}.
		\]
	\end{problem}
	\begin{remark} \label{rmk:change}
		In an instance of signed frustration problem, if we pick an edge $e\in E$, change $c_e$ to $1-c_e$, and $w_e$ to $-w_e$, then we only change the optimal value by a known constant, and leave the optimal assignments unchanged. That is, each instance of maxcut is equivalent (up to an additive constant) to an instance of signed frustration, and vice versa. Moreover, the conversion in both directions can be done in time $O(\max\{|V|,|E|\})$.
	\end{remark}

	\section{Contraction distance and contraction rank}
	\label{sec:parameters}
	
	Suppose $G=(V,E)$ is a graph and $\cB$ is a Eulerian-spanning set of $G$. In this section we will show that $\exc(A_\cB)$, $\rex(A_\cB)$ are independent from the choice of $\cB$, and are equal to the contraction distance and contraction rank of $G$, respectively. 
	Section \ref{contraction} proves the independence and gives the definition of contraction distance and contraction rank. Section \ref{cycle space} characterizes the correspondence between cycle space of a graph and its contracted graph. Section \ref{correspondence} proves the equality between $\exc(A_\cB)$, $\rex(A_\cB)$ and the contraction distance and contraction rank of $G$, and gives FPT algorithms.

	\subsection{Independence} 
	\label{contraction}
	
	We first present the definition of edge contraction and deletion for graphs. They are standard and are presented only for completeness.
	\begin{definition}[edge contraction and deletion]
		Suppose $G=(V,E)$ is a graph. For an edge set $P\subseteq E(G)$, define $G/P$ by identifying all vertices in each connected component of $(V,P)$ and deleting the edges of $P$. Define $G-P$ to be $G-P=(V,E\setminus P)$.
	\end{definition} \par
	It is easy to see that $G/P$ is the graph obtained by
	contracting the edges of $P$ sequentially.

	If $(V_t,E_t)$ is a connected component of $G$, then the projection of $\im M_G$ to $V_t$ is exactly the space of vectors with even Hamming weight. Thus \(\rank(M_G)=|V|-c(G)\) with $c(G)$ being the number of connected components of $G$. And by the rank-nullity theorem, we have 
	\begin{equation}
		\dim\cZ(G)=|E|-|V|+c(G).
		\label{eq:cycle-dimension}
	\end{equation}
	We need a lemma that computes the rank of $(A_\cB)_P$ for any edge set $P$.
	
	\begin{lemma}
		\label{lem:cographic-rank}
		Suppose $G=(V,E)$ is a graph and $\cB$ is a Eulerian basis of $G$. For every $P\subseteq E$,
		\begin{equation}
			\rank((A_{\cB})_P)
			=\dim\cZ(G)-\dim\cZ(G-P)
			=|P|-c(G-P)+c(G).
			\label{eq:cographic-rank}
		\end{equation}
		In particular, this value is independent of $\cB$.
	\end{lemma}
	
	\begin{proof}
		Restricting a Eulerian edge set to the coordinates in $P$ defines a linear map
		\[
		\cZ(G)\longrightarrow\Ftwo^P.
		\]
		Its image is the row space of $(A_{\cB})_P$, and its kernel consists exactly of
		the Eulerian edge sets supported in $E\setminus P$, namely $\cZ(G-P)$.
		Using the rank-nullity theorem and \eqref{eq:cycle-dimension} gives 
		\(\rank((A_{\cB})_P)
		=(|E|-|V|+c(G))
		-(|E|-|P|-|V|+c(G-P))
		=|P|-c(G-P)+c(G).
		\)
	\end{proof}
	
	Let $\mathcal{E}_G$ collect all Eulerian-spanning sets of $G$ and $\tilde{\mathcal{E}}_G$ collect all Eulerian bases of $G$. We are going to prove that the nearly-graphic parameters of $A_\cB$ are independent of the choice of $\cB$.  
	\begin{theorem}
		Suppose $G=(V,E)$ is a graph and $\cB$ is a Eulerian-spanning set of $G$. Then 
		\begin{align*}
			\exc(A_\cB)&=\min_{\cB'\in\mathcal{E}_G}|J_{\cB'}|=\min_{\cB'\in\tilde{\mathcal{E}}_G}|J_{\cB'}|,\\
			\rex(A_\cB)&=\min_{\cB'\in\mathcal{E}_G}\rho_{\cB'}=\min_{\cB'\in\tilde{\mathcal{E}}_G}\rho_{\cB'}.
		\end{align*}
	\end{theorem} \par
	\begin{proof}
		Suppose $A_\cB\in\Ftwo^{r\times E}$. For any $U\in\GL(r,2)$, $UA_\cB$ is the incidence matrix of another Eulerian-spanning set. We have 
		$$\exc(A_\cB)\geq\min_{\cB'\in\mathcal{E}_G}|J_{\cB'}|,\quad \rex(A_\cB)\geq\min_{\cB'\in\mathcal{E}_G}\rho_{\cB'}.$$
		For any Eulerian-spanning set $B'$, there is a Eulerian basis $B''\subseteq B'$, and $|J_{\cB'}|\geq|J_{\cB''}|$, $\rho_{\cB'}\geq\rho_{\cB''}$. Thus
		$$\min_{\cB'\in\mathcal{E}_G}|J_{\cB'}|\ge\min_{\cB'\in\tilde{\mathcal{E}}_G}|J_{\cB'}|, \quad 
		\min_{\cB'\in\mathcal{E}_G}\rho_{\cB'}\ge\min_{\cB'\in\tilde{\mathcal{E}}_G}\rho_{\cB'}.$$
		For any Eulerian basis $\cB'$, there exists a $U\in\GL(r,2)$, such that $UA_{\cB}=\begin{pmatrix} A_{\cB'} \\ 0 \end{pmatrix}$, implying 
		$$\min_{\cB'\in\tilde{\mathcal{E}}_G}|J_{\cB'}|\ge \exc(A_\cB),\quad
		\min_{\cB'\in\tilde{\mathcal{E}}_G}\rho_{\cB'}\geq
		\rex(A_\cB).
		$$
		Hence equality holds all along.
	\end{proof}
	
	For a graph $G=(V,E)$, we set
	\begin{align}
		\cycexc(G)
		&=\min_{\cB\in\tilde{\mathcal{E}}_G}|J_{\cB}|,
		\label{eq:cex}\\
		\rcex(G)
		&=\min_{\cB\in\tilde{\mathcal{E}}_G}\rho_{\cB},
		\label{eq:rcex}\\
		\rho_G(P)&=|P|-c(G-P)+c(G), \text{ for } P\subseteq E, \label{eq:rho-G}\\
		\ctr(G)
		&=\min\{|P|:P\subseteq E,\ G/P\text{ is planar}\},
		\label{eq:ctr}\\
		\rctr(G)
		&=\min\{\rho_G(P):P\subseteq E,\ G/P\text{ is planar}\}.
		\label{eq:rctr}
	\end{align}
	The last three quantities are called the \emph{cographic rank} of $P$, the \emph{contraction distance} of $G$, and the \emph{contraction rank} of $G$,
	respectively.


The inequalities $\rcex(G)\le\cycexc(G)$ and
$\rctr(G)\le\ctr(G)$ are immediate. We note that $\rctr(G)=0$ if and only if $G$ is planar. To see the nontrivial direction, let $P\subseteq E$ satisfy $G/P$ is planar and $\rho_G(P)=0$. The proof of Lemma \ref{lem:cographic-rank} yields that the restriction of
every vector in $\cZ(G)$ to $P$ is zero.  Hence no edge of $P$ lies on a cycle, so every edge of $P$ is a bridge. Construct an auxiliary graph $G_a$, with vertices corresponding to connected components of $G-P$, and edges corresponding to bridges connecting two components. Then the auxiliary graph must be a forest, because a cycle in this auxiliary graph would place every edge of that cycle on a cycle of $G$, contradicting the fact that $\rho_G(P)=0$. Each component of $G-P$ is planar because it is isomorphic to a subgraph of the planar graph $G/P$. Pick any root of $G_a$. Then we can start with planar embeddings of these components and follow the auxiliary tree from its root. We draw each bridge to be both in a face of the father and in the outer face of the child.  This constructs a planar embedding of $G$.

\subsection{Cycle spaces under contraction} \label{cycle space}

\begin{definition}[coordinate projection]
Suppose $G=(V,E)$ is a graph. Let $\pi_P:\Ftwo^{E}\to\Ftwo^{E\setminus P}$ be the natural coordinate projection. The target coordinates are naturally identified with the edges of $G/P$.
\end{definition}

\begin{lemma}[forest contraction witnesses]
\label{lem:forest-witness}
Suppose $G=(V,E)$ is a graph. Let $P\subseteq E$, and let $T$ be a spanning forest of $(V,P)$.  If
$G/P$ is planar, then $G/T$ is planar and $|T|\le|P|$.  Consequently, a minimum-cardinality contraction edge set may be chosen to be a forest.
\end{lemma}

\begin{proof}
Every edge of $P\setminus T$ becomes a loop in $G/T$, and deleting these loops from $G/T$ gives $G/P$. Since adding loops preserves planarity, $G/T$ is planar whenever $G/P$ is planar.  The inequality $|T|\le|P|$ is obvious.
\end{proof}

\begin{lemma}[contraction projection]
\label{lem:contraction-projection}
Let $G=(V,E)$ be a graph. We have, (1) 
for every $P\subseteq E$, the map
\(
\pi_P:\cZ(G)\longrightarrow\cZ(G/P)
\)
is surjective;
(2) if $T\subseteq E$ is a forest, then
\(
\pi_T:\cZ(G)\longrightarrow\cZ(G/T)
\)
is an isomorphism.
\end{lemma}

\begin{proof}
The projection maps Eulerian edge sets to Eulerian edge sets, because the degree parity at a contracted vertex is the sum of the degree parities at the vertices that were identified, with every internal edge contributing twice. \par
Suppose first that $T$ is a forest.  The kernel of $\pi_T$ consists of Eulerian edge sets supported in $T$.  But a forest contains no nonempty Eulerian
edge set, so the kernel is zero. Contracting $T$ decreases both the number of edges and the number of vertices by $|T|$ and leaves the number of connected components unchanged. Thus $\cZ(G)$ and $\cZ(G/T)$ have the same dimension, and $\pi_T$ is an isomorphism. \par
For arbitrary $P$, choose a spanning forest $T$ of $(V,P)$.  In $G/T$, the edges of $P\setminus T$ are loops, and deleting these loops gives $G/P$.  Given $z\in\cZ(G/P)$, extend it to $z'\in\cZ(G/T)$ by assigning value zero to every loop in $P\setminus T$. By the forest case proven above (part 2 of this theorem), there exists a unique $y\in\cZ(G)$ such that $\pi_T(y)=z'$. Therefore $\pi_P(y)=z$ and $\pi_P$ is surjective.
\end{proof}

\subsection{Cardinality and rank correspondences} \label{correspondence}
Before we prove the parameter correspondences, we need a Mac Lane's law of planarity on general graphs. Suppose $G=(V,E)$ is a graph, define the basis number of $G$ to be
\[
b(G)=\min_{\cB\in\tilde{\mathcal{E}}_G}\max_{e\in E(G)} f_{\cB}(e).
\]
We use the convention that the maximum over an empty edge set is zero. By the cycle-decomposition observation described in Remark \ref{cyclebasis}, this definition agrees with the usual
definition in terms of cycle bases. The original Mac Lane's criterion states that a simple graph $G$ is
planar if and only if $b(G)\le2$.  We need and prove the following form. 

\begin{theorem}[Mac Lane's criterion for graphs]
\label{thm:maclane}
A graph $G=(V,E)$ is planar if and only if $b(G)\leq 2$, or equivalently, it has a Eulerian basis $\cB$ satisfying $f_{\cB}(e)\le2$ for every edge $e$.
\end{theorem}

\begin{proof}
First suppose that $G$ is planar. Fix an arbitrary planar embedding of $G$.  For a face $F$ of the embedding, define its \emph{boundary vector} $\partial_2F\in\Ftwo^{E}$ by declaring that $(\partial_2F)_e$ is the number of occurrences of $e$ in the boundary walk of $F$, reduced modulo two.
Then a bridge, which is traversed twice by the same facial boundary walk, has coordinate zero.  For each connected component $K$ of $G$, view the induced drawing of $K$ as a plane embedding in its own right and let $\mathcal F(K)$ be its face set.  Then every cycle $C$ in $K$ corresponds to a Jordan cycle in the plane, and therefore encloses a space. It is easy to see that the sum of all boundaries of faces in $C$ is $C$ itself. Therefore the boundary vectors do span $\cZ(K)$. Omitting one such vector gives a basis, because the only way some of these vectors can sum to zero is that the geometric boundary of the sum of the corresponding faces is empty, meaning that they can only be the whole set of boundary vectors. Taking the union over all components yields a basis of $\cZ(G)$.  Since every edge is incident to at most two faces, it occurs in at most two chosen vectors. 

Conversely, suppose that $G$ has a Eulerian basis $\mc{B}$ in which every edge has
frequency at most two, and let $L_0$ be the set of loops in $G$.  Then the natural coordinate
projection onto $E(G)\setminus L_0$ maps $\cZ(G)$ surjectively onto
$\cZ(G-L_0)$.  Project $\mc{B}$, delete the zero vectors, and select a
basis from the remaining vectors to form a new set $\mc{B}'$. Then each edge can appear in at most two elements in $\mc{B}'$. If a
selected basis element is disconnected, split it into edge-disjoint cycle
and select a basis from the resulting generating family to form a set $\mc{B}''$. Again, each edge can appear in at most two elements in $\mc{B}''$.

Now we subdivide every edge of the resulting loopless graph once.  Then the resulting
graph is simple, and subdivision induces a natural linear cycle-space isomorphism that
preserves the number of appearances of each original edge in a Eulerian basis. Now Mac
Lane's law gives that the subdivided graph is planar.  Suppressing the subdivision vertices and
restoring the loops, and we conclude that the original graph is planar.
\end{proof} \par

\begin{theorem}
\label{thm:equality}
For every graph $G=(V,E)$,
\begin{equation}
	\cycexc(G)=\ctr(G),
	\qquad
	\rcex(G)=\rctr(G).
	\label{eq:equality}
\end{equation}
Moreover, given $P\subseteq E$ such that $G/P$ is planar and a planar embedding of $G/P$, one can construct in polynomial time a Eulerian basis
$\cB$ of $G$ satisfying
\begin{equation}
	J_{\cB}\subseteq P,
	\qquad
	\rho_{\cB}\le\rho_G(P).
	\label{eq:constructive-rank-bound}
\end{equation}
\end{theorem}

\begin{proof}
We first prove 
\begin{equation}
	\ctr(G)\le\cycexc(G),
	\qquad
	\rctr(G)\le\rcex(G).
	\label{eq:first-directions}
\end{equation}
Let $\cB$ be a Eulerian basis and put $P=J_{\cB}$.  By Lemma \ref{lem:contraction-projection}(1), the vectors $\pi_P(B)$, with $B\in\cB$, span $\cZ(G/P)$.  Delete zero vectors and select a basis from the remaining vectors to form a basis $\mc{B}'$.  Every edge of $G/P$ corresponds to an edge outside $P$ and therefore occurs in at most two vectors in $\mc{B}'$. Theorem \ref{thm:maclane} implies that $G/P$ is planar. By Lemma \ref{lem:cographic-rank},
\[
\ctr(G)\le |J_{\cB}|,
\qquad
\rctr(G)\le \rho_G(J_{\cB})
=\rank((A_{\cB})_{J_{\cB}})=\rho_{\cB},
\]
Minimizing the above inequalities over
$\cB$ gives \eqref{eq:first-directions}.

Next we prove
\begin{equation}
	\cycexc(G)\le\ctr(G),
	\qquad
	\rcex(G)\le\rctr(G).
	\label{eq:reverse-directions}
\end{equation}
Let $P\subseteq E(G)$ be such that $G/P$ is planar, and let $T$ be a spanning forest of $(V,P)$.  Lemma \ref{lem:forest-witness} gives that $G/T$ is planar.  By Theorem \ref{thm:maclane}, choose a Eulerian basis $\cB'$ of $G/T$ in which every edge has frequency at most two. Since $T$ is a forest, Lemma \ref{lem:contraction-projection}(2) gives a unique lift $B=\pi_T^{-1}(B')$ of every $B'\in\cB'$. These lifts form a Eulerian basis $\cB$ of $G$. For each edge $e\notin T$, the coordinate is unchanged by the lift, so $f_{\cB}(e)=f_{\cB'}(e)\le2$. Thus
\[
J_{\cB}\subseteq T\subseteq P.
\]
It follows that
\[
|J_{\cB}|\le |P|,
\qquad
\rho_{\cB}
=\rank((A_{\cB})_{J_{\cB}})
\le\rank((A_{\cB})_P)
=\rho_G(P).
\]
Minimizing the first inequality over cardinality-minimal $P$ and the second over rank-minimal $P$ leads to \eqref{eq:reverse-directions}.
Combining \eqref{eq:first-directions} and \eqref{eq:reverse-directions} together proves \eqref{eq:equality}.



For the constructive statement, start from the given embedding of $G/P$. Let $T$ be a spanning forest of $(V,P)$.  Lemma \ref{lem:forest-witness} gives that
$G/T$ is planar. We can compute an embedding of $G/T$ by simply adding the loops corresponding to $P\setminus T$ to $G/P$. By Theorem \ref{thm:maclane}, there is a Eulerian basis $\mc{B}'$ of $G/T$ such that for each edge $e$, $f_{\mc{B}'}(e)\leq 2$. Therefore we only have to compute the lift $\pi_T^{-1}(\mc{B}')$. \par
The unique lift can be computed directly.  Root every tree component of $T$ arbitrarily. For a fixed $B'\in\cB'$, define a vector $x\in\Ftwo^{E}$ by first setting $x_e=B'_e$ for every $e\notin T$.  For a vertex $u\in V$, let $d_x(u)$ be the number of incidences at $u$ among the edges in $(E\setminus T)\cap B'$, with a loop counted twice. For a nonroot vertex $v$ in $T$, let $e_v$ be its parent edge and let $T_v$ be the rooted subtree at $v$. Let $V(T_v)$ be the vertex set of $T_v$. Set
\begin{equation}
	x_{e_v}\equiv\sum_{u\in V(T_v)}d_x(u)\pmod2.
	\label{eq:explicit-lift}
\end{equation}
Let $B=\supp(x)$. Now we show that $B$ is a Eulerian subgraph of $G$. Indeed, for every nonroot vertex, the parity contributed by its parent edge is the same as that of the incidences with $E\setminus T$ and the child edges, so its degree in $\supp(x)$ is even. For a root $r$ with tree component $K$, let $V(K)$ be the vertex set of $K$. Then the sum $\sum_{u\in V(K)}d_x(u)$ equals, modulo two, the degree of $B'$ at the corresponding vertex of $G/T$, which is even. So the root degree is also even. Thus $B=\pi_T^{-1}(B')$ by Lemma \ref{lem:contraction-projection}(2). The construction is polynomial, and the inequalities above give \eqref{eq:constructive-rank-bound}.
\end{proof}

\begin{corollary}
\label{cor:contraction-algorithm}
Suppose $G=(V,E),w,c$ is an instance of Problem \ref{prob:concise}. Moreover, we are given a $P\subseteq E$ such that $G/P$ is planar. Then there are two algorithms finding the optimal value and an optimal assignment in time
\[
O(2^{\rho_G(P)}\max\{|V|,|E|\}^{O(1)})\quad and\quad O(2^{|P|}\max\{|V|,|E|\}^{O(1)})
\]
respectively. The same conclusion holds for the maxcut and signed frustration problems. 
\end{corollary} 

\begin{proof}
By Remark \ref{rmk:loopless}, we assume that $G$ is loopless. A planar embedding of $G/P$ can be found in polynomial time by the planarity-testing algorithm in \cite{HopcroftTarjan1974}.  Construct the Eulerian basis $\cB$ by Theorem \ref{thm:equality}. Then $(I,P)$ is a nearly-graphic certificate for $A_{\cB}$. so Theorem \ref{thm:affine} gives the stated running time.  The special cases follow from Corollary \ref{cor:maxcut} and Remark \ref{rmk:change}.
\end{proof} \par
Let $n=|V|,m=|E|$. If we relax the exponent above 2 to $|P|$, we can actually apply a more delicate algorithm that takes time $O(2^{|P|}(\max\{m,n\}^{3/2}\log(\max\{m,n\})+|P|\max\{m,n\}))$. 
\begin{theorem}\label{thm:faster}
Suppose $G=(V,E),w,c$ is an instance of Problem \ref{prob:concise}. Moreover, we are given $P\subseteq E(G)$ such that $G/P$ is planar. Let $n=|V|,m=|E|$. Then the optimal value and an optimal assignment can be found in time
$$O(2^{|P|}(\max\{m,n\}^{3/2}\log(\max\{m,n\})+|P|\max\{m,n\})).$$
The same conclusion holds for maxcut and signed frustration.
\end{theorem}
This result can be applied to the parameterization by crossing numbers. For the proof and the application, see Appendix \ref{ab}.

\begin{proof}[Proof of Theorem \ref{thm:intrinsic-contraction-fpt}]
Let $N=\max\{|V|,|E|\}$. The Planar Contraction decision problem asks whether an input pair $(G,k)$ admits a set $P\subseteq E(G)$ with $|P|\le k$ such that $G/P$ is planar. There is an $f(k)N^{c}$-time algorithm for solving it \cite{GolovachEtAl2013,AbelloEtAl2017}. Put $F(k)=\max_{0\le i\le k}f(i)$.  Run the decision algorithm for $i=0,1,\ldots$ until the first time that the algorithm accepts. This process terminates, because contracting a spanning forest of $G$ produces a planar graph consisting only of loops. If $k=\ctr(G)$ is the first value for which the algorithm accepts, the total running time of these calls is at most $(k+1)F(k)N^c$. \par
Now we recover the set $P$ by self-reduction. Call $(G,k)$ a positive instance if there is a set $P\subseteq E(G)$ with $|P|\le k$ such that $G/P$ is planar. If a nonplanar instance $(G,k)$ is positive, then there must be an edge $e$ such that the instance $(G/e,k-1)$ is positive. Therefore we go through all edges, choose one $e$ such that the instance $(G/e,k-1)$ is positive, contract it, and repeat the procedure until the graph is planar. This uses at most $k|E|$ decision calls. The search therefore takes at most $k|E|F(k)N^c$ time.  For the returned set $P$, one has $\rho_G(P)\le |P|\le k$. Applying Corollary \ref{cor:contraction-algorithm} concludes the proof.
\end{proof} \par

\section{Conclusion and outlook}
In this paper, we give an FPT algorithm solving BAO when a nearly-graphic certificate is given, which is parameterized by the support size and non-graphic rank. We explore the application of this algorithm to Max-Lin$_2$. When the constraints of Max-Lin$_2$ are defined by a graph, we use Eulerian-spanning set to describe the converted affine system, and prove that the nearly-graphic parameters of $A_{\cB}$, i.e. $\cycexc(G)$, $\rcex(G)$ are equal to $\ctr(G)$, $\rctr(G)$, respectively. We therefore give FPT algorithms solving Max-Lin$_2$ on graphs. We specifically obtain an FPT algorithm solving Max-Lin$_2$ on graphs, which is parameterized by contraction distance and needs no information other than the problem itself. Moreover, we prove an arbitrarily large separation between $\ctr(G)$ and $\rctr(G)$. \par
Several problems are left open for the future. Finding a nearly-graphic certificate $(U,J)$ that has optimal, or nearly optimal parameters for the affine system $Bx=d$ is an important one. Additionally, one may try to minimize the exponents above $\max\{|V|,|E|\}$ in algorithms present in Section \ref{sec:maxlingraph}, which we do not give explicitly.

\section*{Acknowledgement}
This research was supported by the National Natural Science Foundation of China (Nos. 12325112, 12288101).

\appendix
\section{Proof of Lemma \ref{lem3main}} \label{aa}
Now we are given a graph $G=(V,E)$ without loops, a vertex set $V_t\subseteq V$. Our goal is to find the minimum weighted set $E'\subseteq E$ such that $\partial_G(E')=V_t$, or claim that no such $E'$ exists. If $G$ is not connected, we can solve the problem on each connected component of $G$, respectively. Therefore, we assume that $G$ is connected from now on. Moreover, if $2\nmid |V_t|$, then by a simple counting argument there can be no such $E'$. Therefore, we assume that $2\mid |V_t|$ from now on. \par
We first convert the problem to a maximum weighted Eulerian subgraph problem. Let $V_o\subseteq V$ be the set of vertices of odd degree in $G$, and $V_c=V_o\triangle V_t$. Now we need to compute an edge set $E_c\subseteq E$ such that $\partial_G(E_c)=V_c$. Specifically, we have the following lemma:
\begin{lemma}
Let $G=(V,E)$ be a connected graph, and $Q\subseteq V$ satisfy $2\mid|Q|$. Then we can find an edge set $E_Q\subseteq E$ such that $\partial_G(E_Q)=Q$ in time $O(\max\{|V|,|E|\})$.
\end{lemma}
\begin{proof}
First, we compute a rooted spanning tree $T=(V,E_T)$ of $G$ in time $O(\max\{|V|,|E|\})$. Let the root be $r$. We pick the edge set $E_Q\subseteq E_T$ as follows. Let $e=(u,v)\in E_T$, and suppose $v$ is the son of $u$. Denote by $T_v$ the subtree rooted at $v$, and $V(T_v)$ the set of vertices in $T_v$. Then $e\in E_Q$ if and only if $|V(T_v)\cap Q|$ is odd. The set $E_Q$ can be easily computed in time $O(\max\{|V|,|E|\})$ by tracing from the leaves to the root. Therefore we only need to verify that $\partial_G(E_Q)=Q$. Now, for each non-root vertex $v$, the degree of $v$ in the subgraph $(V,E_Q)$ is equal (modulo 2) to $|V(T_v)\cap Q|-\sum_{w\ is\ a\ son\ of\ v}|V(T_w)\cap Q|$, which is 1 if and only if $v\in Q$. Because there is only one root $r$, the parity condition forces that $r\in Q\Leftrightarrow r\in \partial_G(E_Q)$. Therefore $\partial_G(E_Q)=Q$.
\end{proof}
By the above lemma, we can compute the set $E_c\subseteq E$ such that $\partial_G(E_c)=V_c$ in time $O(\max\{|V|,|E|\})$. Let $W$ be the sum of the absolute values of weights of $E$. For every $e\in E_c$, we add a duplicate of $e$ to the graph $G$, and weight the duplicated edge by $W+1$. Denote the graph obtained by $G_{aug}=(V,E_{aug})$, then it suffices to find $E_1'\subseteq E_{aug}$ with minimum total weight such that $\partial_{G_{aug}}(E_1')=V_t$. Moreover, $V_t$ is the set of odd vertices in $G_{aug}$. Therefore, $\forall E_1'\subseteq E_{aug},\partial_{G_{aug}}(E_1')=V_t$, we have that $(V,E_{aug}\rs E_1')$ is a Eulerian subgraph, and vice versa. Therefore, it suffices to find a maximum weighted Eulerian subgraph of $G_{aug}$. We note that $|E_{aug}|\leq 2|E|$ because $E_c\subseteq E$. Moreover, if $G$ is planar, then $G_{aug}$ is also planar, because adding duplicate edges does not affect planarity. Obviously a planar embedding of $G_{aug}$ can be computed in time $O(\max\{m,n\})$.\par
Now we convert the maximum weighted Eulerian subgraph problem on $G_{aug}$ to a maximum weighted perfect matching problem, using a procedure almost identical to that in \cite{113}. Starting with the graph $G_{aug}$, the conversion consists of two steps:\par
\begin{figure}[h]
\centering 
\includegraphics[height=135pt]{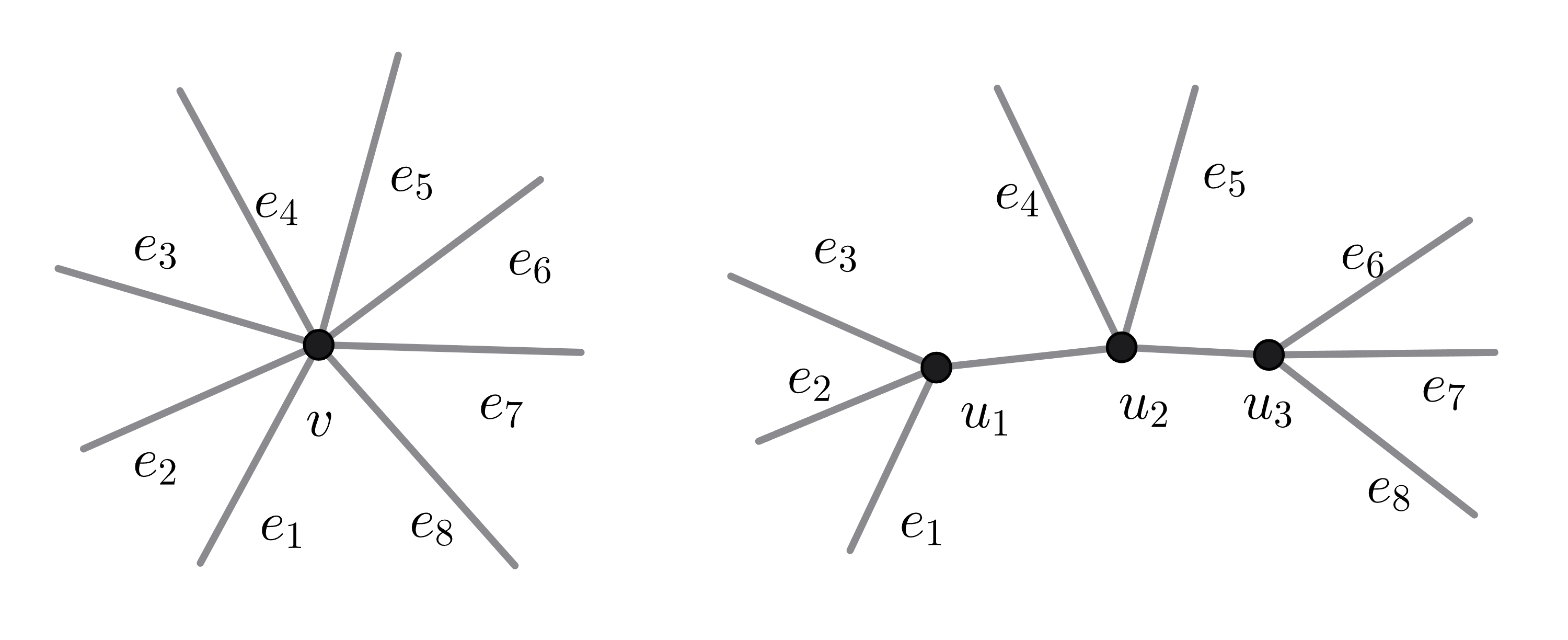}
\caption{An example of modification in step 1. The vertex $v$ is of degree 8, and is split into three vertices.}
\label{f4}
\end{figure}

Step 1. For every vertex $v\in V$ of degree at least five in $G_{aug}$, do the following: let the edges incident to $v$ be $e_1,e_2,\cdots,e_r$, in cyclic order. Split $v$ into $q=\lfloor \frac{r-1}{2}\rfloor$ vertices, $u_1,\cdots,u_q$. Add a path $u_1u_2\cdots u_q$, and weight all edges on the path by zero. Attach $e_1,e_2,e_3$ to $u_1$. Attach $e_{2i},e_{2i+1}$ to $u_i$, for $i=2,3,\cdots,q-1$. Attach $e_{2q},e_{2q+1},\cdots,e_r$ to $u_q$. This step of modification is depicted in Figure \ref{f4}. \par
Let the graph obtained after step 1 be $G^{(1)}=(V^{(1)},E^{(1)})$, which is still planar if $G_{aug}$ is planar, as is shown in Figure \ref{f4}. Moreover, every vertex in $G^{(1)}$ is of degree at most 4. Every Eulerian subgraph of $G^{(1)}$ corresponds to a Eulerian subgraph of $G_{aug}$ with the same total weight, and vice versa. Therefore we only need to find a maximum weighted Eulerian subgraph in $G^{(1)}$. We note that $|V^{(1)}|\leq |V|+|E|\leq\max\{m,n\}+m=O(\max\{m,n\})$,$|E^{(1)}|\leq|E|+|V^{(1)}|=O(\max\{m,n\})$.\par
Step 2. For every vertex $v$ in $G^{(1)}$, do one of the following according to the degree of $v$:\par
(i) If $v$ is of degree one, let the edge incident to $v$ be $e_1$. Split $v$ into $K_2$, weight the edge in $K_2$ by zero, and attach $e_1$ to a vertex in $K_2$. \par
(ii) If $v$ is of degree two, let the edges incident to $v$ be $e_1,e_2$. Split $v$ into $K_2$, weight the edge in $K_2$ by zero, and attach $e_1,e_2$ to different vertices in $K_2$. \par
(iii) If $v$ is of degree three, let the edges incident to $v$ be $e_1,e_2,e_3$. Split $v$ into $K_4$, weight the edges in $K_4$ by zero, and attach $e_1,e_2,e_3$ to three different vertices in $K_4$. \par
(iv) If $v$ is of degree four, let the edges incident to $v$ be $e_1,e_2,e_3,e_4$. Split $v$ into $K_4$, weight the edges in $K_4$ by zero, and attach $e_1,e_2,e_3,e_4$ to four different vertices in $K_4$. \par
\begin{figure}[h]
\centering 
\includegraphics[height=135pt]{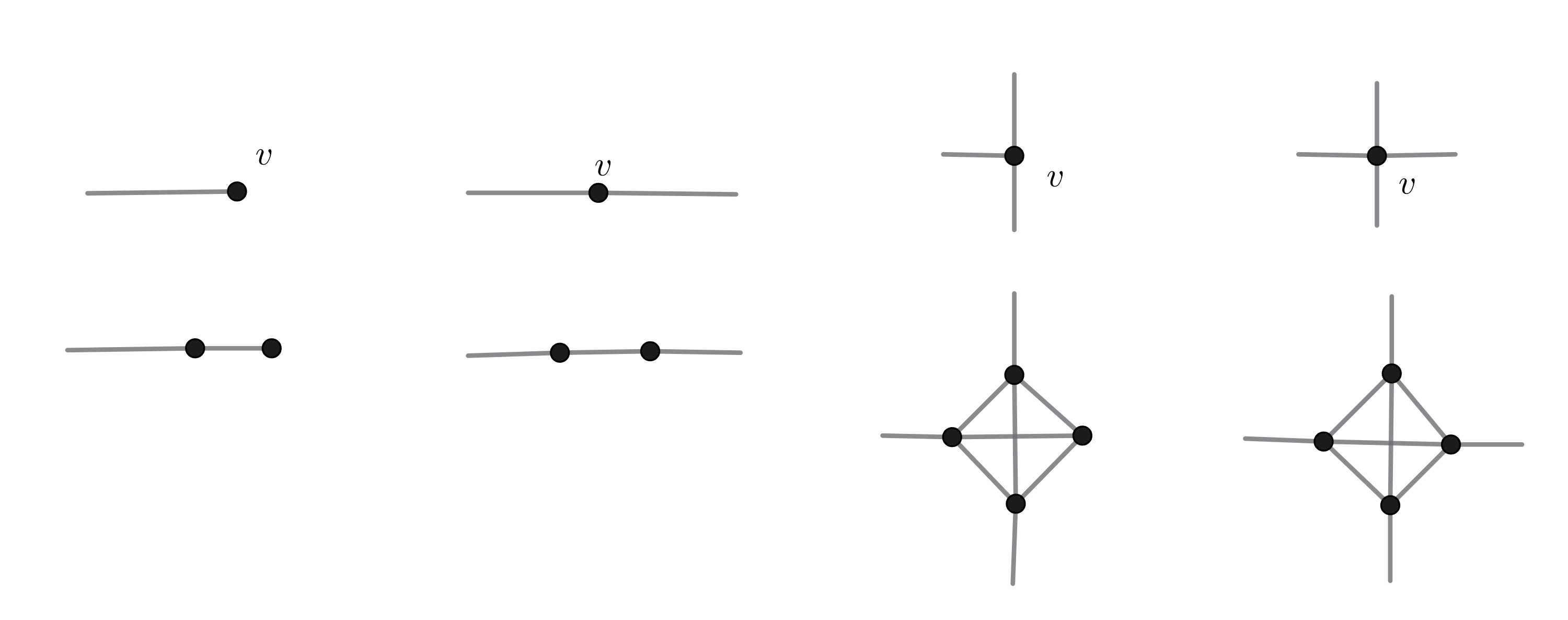}
\caption{The modification in step 2. The figures, from left to right, correspond to the cases where $v$ is of degree 1,2,3,4, respectively.}
\label{f5}
\end{figure}

The modification in this step is depicted in Figure \ref{f5}. \par
Let $G^{(2)}=(V^{(2)},E^{(2)})$ be the graph obtained after this step. Then $G^{(2)}$ is actually a simple graph. It can be verified that, for every Eulerian subgraph of $G^{(1)}$, its corresponding edges in $G^{(2)}$ is a subset of a perfect matching with the same total weight. And for every perfect matching in $G^{(2)}$, the edges in $G^{(1)}$ corresponding to matching edges in $G^{(2)}$ form a Eulerian subgraph with the same total weight. The verifications are similar to those in \cite{113}, which already cover the cases where $v$ is of degree 3 and 4. Therefore we omit the verifications here. We note that $|V^{(2)}|\leq 4|V^{(1)}|=O(\max\{m,n\})$,$|E^{(2)}|\leq|E^{(1)}|+6|V^{(1)}|=O(\max\{m,n\})$. \par
Now we only need to find a maximum weighted perfect matching in $G^{(2)}$. First, by adding a suitable large number to all weights, we can convert the problem to finding a maximum weighted matching (note that there exists a perfect matching in $G^{(2)}$, because empty graph is a Eulerian subgraph of $G^{(1)}$). In general case, we can simply apply the Algorithm in \cite{110} to solve the problem in time $O((\max\{m,n\})^3)$. If $G$ is planar, then $G_{aug}$ is planar as well. We can exploit the planarity to reduce the time complexity. Now we describe an algorithm similar to that in \cite{113,114}, using some results of this work:
\begin{theorem}[\it{Theorem 1 of \cite{114}}]\label{thm62}
Let $G=(V,E)$ be any undirected simple planar graph of $n$ vertices. Each vertex is assigned a nonnegative weight, such that all weights sums to one. Let $n=|V|$. Then there is an $O(n)$-time algorithm finding a partition of $V$ into three sets $A,B,C$ satisfying the following conditions. Neither $A$ or $B$ has total weight larger than $2/3$. $|C|\leq 2\sqrt{2n}$. No edge of $E$ joins a vertex in $A$ and a vertex in $B$.
\end{theorem} \par
In the algorithm, we need to apply the above algorithm to the graph $G^{(1)}$, and its induced subgraphs. Although $G^{(1)}$ itself is not simple, each vertex has degree at most 4, and the same condition holds for its induced subgraphs. Therefore we can modify the algorithm and use the following corollary.
\begin{corollary}\label{col61}
Let $G=(V,E)$ be any undirected planar graph, such that each vertex has degree at most 4. Each vertex is assigned a nonnegative weight, such that all weights sums to one. Let $n=|V|$. Then there is an $O(n)$-time algorithm finding a partition of $V$ into three sets $A,B,C$ satisfying the following conditions. Neither $A$ or $B$ has total weight larger than $2/3$. $|C|\leq 2\sqrt{2n}$. No edge of $E$ joins a vertex in $A$ and a vertex in $B$.
\end{corollary} \par
\begin{proof}
By the degree condition, $|E|\leq 2|V|=2n$. We compute the underlying simple graph in time $O(n)$, and construct the separator $A,B,C$ for the simple graph as in Theorem \ref{thm62}. Therefore no edge joins a vertex in $A$ and a vertex in $B$ in the original graph.
\end{proof}
\begin{theorem}[\it{see Lemma 2 in \cite{114} and the comments below it}]\label{thm63}
Let $G=(V,E)$ be an undirected simple graph with edges weighted, and $v\in V$. Let $n=|V|,m=|E|$. Suppose we know the maximum weighted matching of $G-v$, then a maximum weighted matching of $G$ can be computed in time $O(m\log n)$.
\end{theorem} \par
Now we describe the algorithm, which begins with the simple graph $G^{(2)}$: \par
Step 1. If $G=(V(G),E(G))$ contains at most one vertex, return the empty set as the maximum weighted matching. \par
Step 2. Let the vertices of $G^{(1)}$ corresponding to vertices in $G$ form a set $V$. For each vertex $v\in V$, let $\sigma_v\in\{2,4\}$ be the number of vertices in the gadget of $G^{(2)}$ that replaces $v$. Assign $v$ the normalized weight $\frac{\sigma_v}{\sum_{v\in V}\sigma_v}$. Apply the algorithm of Corollary \ref{col61} on the induced graph $G^{(1)}[V]$ to get a partition of $V$ into three sets $A,B,C$, where $|C|=O(\sqrt{|V|})$, the total weight of each of $A,B$ is at most $2/3$, and no edge of $G^{(1)}[V]$ joins a vertex in $A$ and a vertex in $B$. Let $A_p,B_p,C_p\subseteq V(G)$ be the unions of gadgets corresponding to $A,B,C$. Then $|C_p|\leq4|C|=O(\sqrt{|V|})=O(\sqrt{|V(G)|}), |A_p|\leq 2|V(G)|/3,|B_p|\leq 2|V(G)|/3$, and no edge of $G$ joins a vertex in $A_p$ and a vertex in $B_p$. Apply our algorithm recursively to $G[A_p],G[B_p]$ to find the maximum weighted matching in $G[A_p],G[B_p]$. \par
Step 3. Add vertices in $C_p$ to $G[A_p\cup B_p]$ one at a time. Each time a vertex is added, apply Theorem \ref{thm63} to find a maximum weighted matching in the new graph. \par
The description is completed. \par
In Step 2, there are no edges of $G$ joining a vertex in $A_p$ and a vertex in $B_p$, therefore the union of maximum weighted matchings of $G[A_p],G[B_p]$ is indeed the maximum weighted matching of $G[A_p\cup B_p]$. Therefore the algorithm is correct. Let $t(n)$ be the running time of this algorithm on a graph with at most $n$ vertices. Then
\begin{equation}
\begin{aligned}
	& t(1)=c_1 \\
	\forall n>1,\ &t(n)\leq\max\{t(n_1)+t(n_2)+c_2n^{3/2}\log n:n_1+n_2=n,n_1,n_2\leq 2n/3\}.
\end{aligned}
\end{equation}
Where $c_1,c_2$ are positive constants. An inductive argument shows that $t(n)=O(n^{3/2}\log n)$. Therefore the time needed to compute a maximum weighted matching in $G^{(2)}$ is \\ $O(\max\{m,n\}^{3/2}\log(\max\{m,n\}))$. \par
Moreover, converting the problem to a maximum weighted matching problem costs time at most $O(\max\{m,n\})$. Solving the maximum weighted matching problem costs time $O(\max\{m,n\}^{3/2}\log(\max\{m,n\}))$, and recovering the answer $E'$ based on the solution also costs time at most $O(\max\{m,n\})$. Therefore the procedure in this appendix costs an overall $O(\max\{m,n\}^{3/2}\log(\max\{m,n\}))$ time.

\section{Proof of Theorem \ref{thm:faster}, and the application}\label{ab}
\subsection{Proof of Theorem \ref{thm:faster}, and a corollary}
\begin{proof}[Proof of Theorem \ref{thm:faster}]
We assume that $G$ is loopless by Remark \ref{rmk:loopless}. \par
Pick $T\subseteq P$ such that $(V,T)$ is a spanning forest of $(V,P)$. Then $G/T$ is also planar. Pick any planar embedding of $G/T$ (which can be computed in time $O(\max\{m,n\})$), and take the Eulerian-spanning set $\cB_T$ to be the set of boundary vectors of all faces in the embedding. Let $\cB=\pi_T^{-1}(\cB_T)$, where $\pi_T$ is as in Lemma \ref{lem:contraction-projection}. Then $\cB$ is a Eulerian-spanning set of $G$. Note that the size of the matrix $A_{\cB}$ may already exceed the expected running time. But we can store it in a clever way. Specifically, we simply store the support of each column of $A_{\cB}$. \par
Now we consider the cost of storing the support of each column of $A_{\cB}$. By Theorem \ref{thm:equality}, $|\cB|=|\cB_T|$, and $|\cB_T|$ is simply the number of faces of $G/T$, which cannot exceed $O(\max\{m,n\})$ by Euler's formula. Now we consider the total cardinality of edge sets in $\cB$. This value is equal to the number of 1's in $A_{\cB}$, which is in turn equal to 
$$\sum_{e\in E}f_{\cB}(e).$$
For $e\notin T$, $f_{\cB}(e)\le2$. For $e\in T$, $f_{\cB}(e)\le|\cB|=O(\max\{m,n\})$. Therefore the total cardinality of edge sets in $\cB$ is $O((|T|+1)\max\{m,n\})$. \par
By \eqref{eq:graph-min}, we only need to solve the following optimization problem and compute an optimal assignment:
$$\min\{w^{\mathsf T}z:A_{\cB}z=A_{\cB}c,\ z\in\Ftwo^E\}.$$
Note that we can compute $A_{\cB}c$ again by simply determining the value for each edge one by one, which takes a total of $O((|T|+1)\max\{m,n\})$ time. Our algorithm is as follows: for every $z_T\in \Ftwo^T$, compute
$$Q(z_T)\triangleq\min\{w_{\overline T}^{\mathsf T}z_{\overline T}:(A_{\cB})_{\overline T}z_{\overline T}=A_{\cB}c+(A_{\cB})_Tz_T,\ z_{\overline T}\in\Ftwo^{\overline T}\}.$$
We can again compute $(A_{\cB})_Tz_T$ by simply determining the value for each edge one by one, which takes a total of $O((|T|+1)\max\{m,n\})$ time. Let $G_d=(V_d,E_d)$ be the planar dual of $G/T$, which is again planar. Then there is a one-to-one correspondence between $V_d$ and $\cB$. Therefore we can also write $G_d$ as $G_d=(\cB,E_d)$. We have $|E_d|\leq |E|=m$ by planar duality. Each edge $e\in E\setminus T$ corresponds to an edge $e'=(B_i,B_j)\in E_d$ such that the boundary vectors of the two faces incident to $e$ are $B_i,B_j$. Therefore $(A_\cB)_{\overline T}=M_{G_d}$. Then $\forall z\in\Ftwo^E, (A_{\cB})_{\overline T}z_{\overline T}=A_{\cB}c+(A_{\cB})_Tz_T$ is equivalent to $\partial_{G_d}(\supp(z_{\overline T}))=\supp(A_{\cB}c+(A_{\cB})_Tz_T)$. After picking the loops with negative weight and removing all loops, we can use Lemma \ref{lem3main} to solve the optimization problem and recover an optimal assignment in time $O(\max\{m,n\}^{3/2}\log \max\{m,n\})$.\par
Last, we compute $\min_{z_T}\{Q(z_T)+w_{T}^{\mathsf T} z_T\}$ and get the optimal value and assignment. The total running time is therefore
$$
\begin{aligned}
	&O(2^{|T|}(\max\{m,n\}^{3/2}\log(\max\{m,n\})+|T|\max\{m,n\}) \\
	=&O(2^{|P|}(\max\{m,n\}^{3/2}\log(\max\{m,n\})+|P|\max\{m,n\}).
\end{aligned}
$$
\end{proof} \par
Because we can also add zero-weighted edges (and arbitrarily assign the violation value) without affecting the optimal solution and assignment of the problem. Then we have the following:
\begin{corollary} \label{col:faster}
Suppose $G=(V,E),w,c$ is an instance of Problem \ref{prob:concise}. Moreover, we are given another edge set
$P$, such that if we denote $G_{aug}=(V,E\cup P)$, then $G_{aug}/P$ is planar. Let $n=|V|,m=|E|,k=|P|,N=\max\{m,n\}$. Then an optimal assignment can be found
in
$$O(2^k((N+k)^{3/2}\log(N+k) + k(N+k)))$$
time.  The same conclusion holds for maxcut and signed frustration.
\end{corollary}

\subsection{Application to the case of crossing numbers}
We first present the definition of $s$-planar graphs \cite{111} and crossing number of a given embedding.
\begin{definition}
A graph $G=(V,E)$ is $s$-planar if it can be embedded in the plane such that each edge is crossed at most $s$ times. The crossing number of an embedding is obtained in the following way: count the crossings separately for each pair of edges, and for each crossing point of the pair of edges. 
We say $G$ has an $s$-planar embedding of crossing number $k$, if $G$ can be embedded in the plane such that each edge is crossed at most $s$ times, and the crossing number of this embedding is $k$. 	We define the crossing number of $G$ to be the minimum crossing number among all embeddings of $G$ in the plane.
\end{definition}
\begin{remark}
Given an embedding of $G$ in the plane, if there are at least three edges that intersect in one point, we can adjust the edges a little in a local way such that each pair of edges intersect in a different point, without changing the total number of crossings. Therefore, we will always assume that $G$ is drawn in the plane so that situations in which three edges intersect in one point do not occur.
\end{remark}
\begin{lemma}\label{lem71}
Suppose $G=(V,E)$ is a simple graph that has a 2-planar embedding of crossing number $k$. Then there is another edge set $E_1$ such that $|E_1|\leq k$, and if we let $G_{aug}=(V,E\cup E_1)$, then the underlying simple graph of $G_{aug}/E_1$, which we denote by $G_p$, is planar. Moreover, we can compute $E_1$ and the planar embedding of $G_p$ in time $O(km)$.
\end{lemma}
\begin{proof}
We construct a procedure consisting of at most $k$ rounds. In each round, we do one of the following three things:\par
i. Re-embed (or redraw) $G$ in the plane. \par
ii. Contract an edge in $G$ while merging multiple edges and removing loops. \par
iii. Pick an edge not in $G$, add it to $G$, and contract it while merging multiple edges and removing loops. \par
We will show that after each round, $G$ is a simple graph that has a 2-planar embedding of smaller crossing number. Therefore, after at most $k$ rounds, we will get a planar graph, and the edges we have contracted during the procedure certainly form the set $E_1$ we wanted. \par
Now suppose $G=(V,E)$ is a simple graph with a 2-planar embedding of crossing number $r$. Pick any intersection point $x$, and suppose it is the intersection of $e_1$ and $e_2$. We distinguish between three cases:\par
(1) If $e_1$ and $e_2$ intersect at least twice, then by 2-planarity of the embedding, they must intersect exactly twice, and cannot possibly intersect any other edge. Therefore, $e_1,e_2$ can be redrawn in the way depicted in Figure \ref{f1}, such that $e_1$ and $e_2$ do not cross each other, and do not cross any other edges. Therefore the redrawing is indeed 2-planar, and the crossing number is reduced.\par
(2) If $e_1$ and $e_2$ intersect exactly once at $x$, and share an endpoint, say, $v_1$. Let $e_1=(v_1,v_2),e_2=(v_1,v_3)$. By 2-planarity, the segments $v_1x(\text{on }e_1),v_1x(\text{on }e_2),xv_2,xv_3$ can intersect other edges at most once each. Therefore, if we redraw the edges $e_1,e_2$ in the way depicted in Figure \ref{f1}, then they do not intersect each other, and can only intersect at most twice with other edges each. Therefore the redrawing is indeed 2-planar. Moreover, the crossing number is also reduced after the redrawing.
\begin{figure}[h]
	\centering 
	\includegraphics[height=150pt]{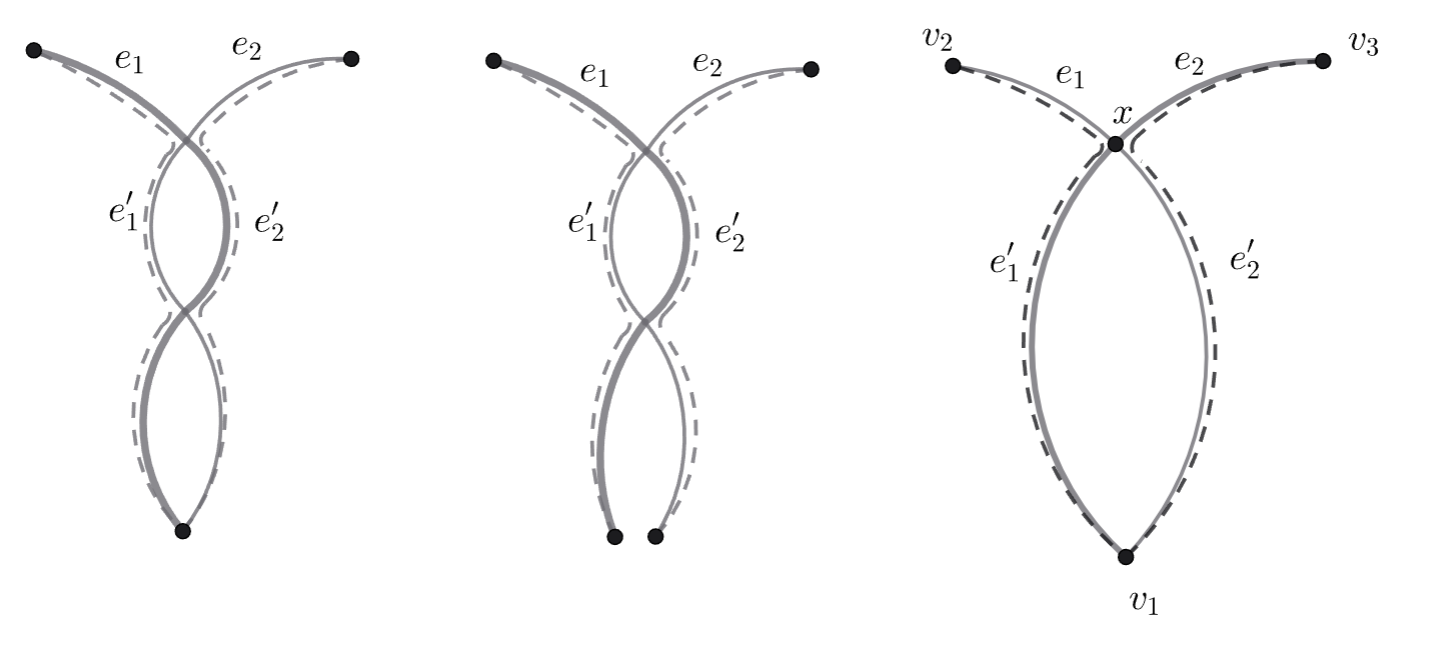}
	\caption{the case of two edges intersecting two times (figures in the left and middle, we distinguish cases whether they share an endpoint or not), and the case where two edges share an endpoint, but intersect only once (figure in the right). The original edges are those in solid lines, while the redrawn ones are those in dotted lines.}
	\label{f1}
\end{figure} \par
(3) If the endpoints of $e_1$ and $e_2$ are pairwise different, and they intersect only once on $x$. Let $e_1=(v_1,v_2),e_2=(v_3,v_4)$. By 2-planarity of the embedding, the segments $xv_1,xv_2$ on $e_1$ can only be crossed once by other edges altogether. Therefore there must be a segment, say $xv_1$, that is not crossed by any other edge. Similarly, let $xv_3$ be the segment on $e_2$ that is not crossed by any other edge. Let $e=(v_1,v_3)$. We further distinguish between two cases: \par
(i) If $e\in E$, we first contract $e$ in $G$ while preserving multiple edges and loops, and redraw the edges $e_1,e_2$ and all edges incident to $v_1,v_3$ in the way depicted in Figure \ref{f2}. Specifically, we remove the edge $e$, move the vertices $v_1,v_3$ to $x$ along the segments $v_1x,v_3x$, and extend the edges incident to $v_1,v_3$ along the corresponding segments. Then we adjust the segments $v_1x,v_3x$ on edges incident to $v_1,v_3$ in a local way so that they do not cross each other, and do not cross other edges as well.\par
Then an intersection point in the redrawing must also be an intersection point of $G$. Moreover, $x$ is no longer an intersection point of the redrawing. Therefore the redrawing is 2-planar, and of smaller crossing number. Then we merge multiple edges and remove loops, which cannot affect 2-planarity or increase the crossing number.\par
(ii) If $e\notin E$, we add the edge $e$ to $G$ (we note that the way we draw the edge $e$ on the plane is irrelevant, because we will contract it soon). Then we contract $e$ while preserving multiple edges and loops, and redraw the edges $e_1,e_2$ and all edges incident to $v_1,v_3$ in the way depicted in Figure \ref{f3}. Specifically, we remove the edge $e$, move the vertices $v_1,v_3$ to $x$ along the segments $v_1x,v_3x$, and extend the edges incident to $v_1,v_3$ along the corresponding segments. Then we adjust the segments $v_1x,v_3x$ on edges incident to $v_1,v_3$ in a local way so that they do not cross each other, and do not cross other edges as well.\par
Then an intersection point in the redrawing must also be an intersection point of $G$. Moreover, $x$ is no longer an intersection point of the redrawing. Therefore the redrawing is indeed 2-planar, and of smaller crossing number. Then we merge multiple edges and remove loops, which cannot affect 2-planarity or increase the crossing number.\par
\begin{figure}[h]
	\centering 
	\includegraphics[height=135pt]{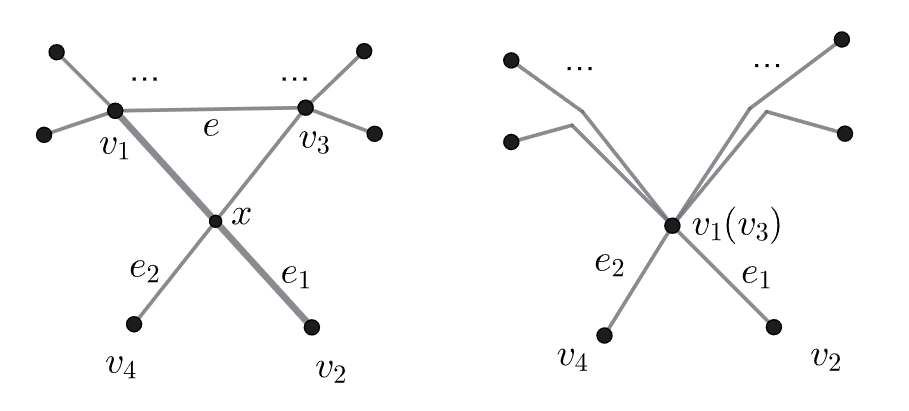}
	\caption{The case where the endpoints of $e_1$ and $e_2$ are pairwise different, and they intersect only once on $x$. If $e=(v_1,v_3)\in E$, we contract $e$ to the intersection point $x$, and redraw all edges incident to $v_1,v_3$ as depicted above (we have not merged multiple edges or removed loops so far).}
	\label{f2}
\end{figure} \par
\begin{figure}[h]
	\centering 
	\includegraphics[height=135pt]{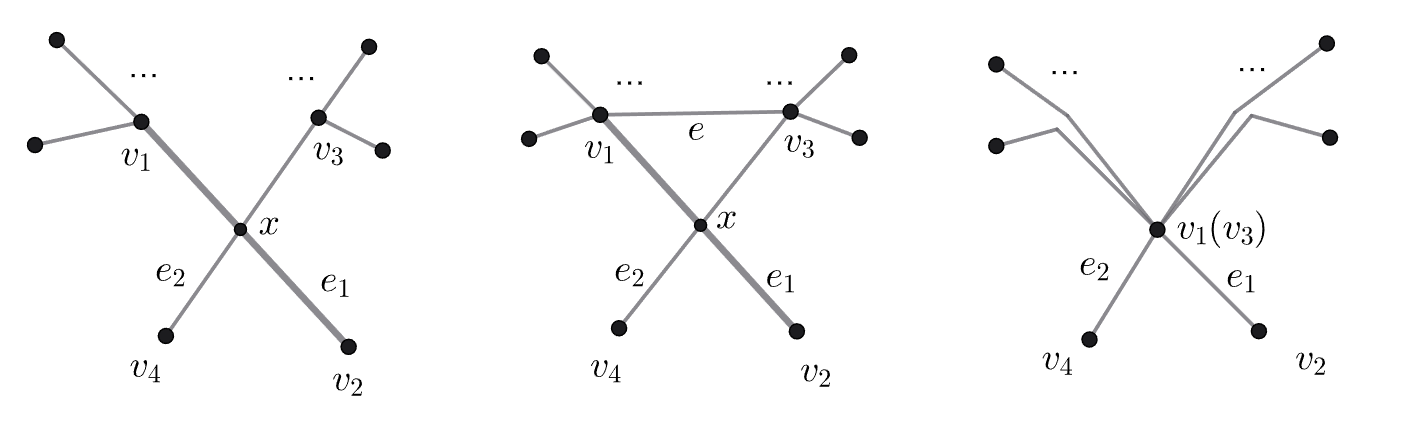}
	\caption{The case where the endpoints of $e_1$ and $e_2$ are pairwise different, and they intersect only once on $x$. If $e=(v_1,v_3)\notin E$, we first add $e$ to $G$, then contract $e$ to the intersection point $x$, and redraw all edges incident to $v_1,v_3$ as depicted above (we have not merged multiple edges or removed loops so far).}
	\label{f3}
\end{figure} \par
Now we consider the computational complexity of finding $E_1$ and planar embedding of $G_p$. Each round of our procedure can be done in time $O(m)$, since there are at most $m$ edges to be redrawn. Therefore the total running time of the procedure is $O(km)$.
\end{proof}
\begin{theorem}\label{thm71}
Suppose $G=(V,E), w, c$ is an instance of Problem \ref{prob:concise}, where \(G\) is simple. Moreover, suppose that we are given a $2$-planar embedding of \(G\) with crossing number \(k\). Let \(n=|V|\) and \(m=|E|\). Then there is an algorithm running in \(O(2^k(n^{3/2}\log n+kn))\) time that solves the problem and computes an optimal assignment.	
\end{theorem}
\begin{proof}
By \cite{111}, $m\leq 5n-10=O(n)$. Each edge is crossed at most twice, therefore $k\leq 2m=O(n)$. Apply Lemma \ref{lem71} and Corollary \ref{col:faster}.
\end{proof} \par
\begin{theorem}\label{thm72}
Suppose $G=(V,E),w,c$ is an instance of Problem \ref{prob:concise}, and $G$ is simple. Moreover, we are given a planar embedding of $G=(V,E)$ with crossing number $k$. Let $n=|V|,m=|E|$. Then there is an $O(2^k((n+k)^{3/2}\log (n+k)+(n+k)k))$-time algorithm solving the problem, and computing the optimal assignment. 
\end{theorem}
\begin{proof}
By Remark \ref{rmk:change}, we can assume $\forall e\in E, w_e\geq 0$. \par
For any edge $e=(v_1,v_2)\in E$ do the following: let a point move from $v_1$ to $v_2$ through edge $e$, and the intersection points it encounters be $u_1,\cdots,u_l$ in chronological order. If $l<3$, we do nothing. Otherwise, for each $i\in\{1,2,\cdots,\lfloor\frac{l-1}{2}\rfloor\}$, add two vertices $z_{2i-1},z_{2i}$ between $u_{2i},u_{2i+1}$ on $e$. Weight every edge on the path $v_1z_1z_2\cdots z_{2\lfloor\frac{l-1}{2}\rfloor}v_2$ by $w_e$, and set the coordinates of $c$ on these edges to $c_e$. Let the resulting graph be $G_2$. Then $G_2$ is a 2-planar graph with at most $n+2k$ vertices. Suppose the weight and violation vectors of $G_2$ are $w',c'$. Then the instance $(G_2,w',c')$ is equivalent to that of $G$ in the following sense: \par
View a vector $x\in\Ftwo^V$ as a vertex partition. For each subdivided edge $e\in E$, distinguish between two cases: \par
(1) $c_e=1$ \par
If $v_1$ and $v_2$ belong to the same set of the partition, then at most $2\lfloor\frac{l-1}{2}\rfloor$ edges on the path $v_1z_1z_2\cdots z_{2\lfloor\frac{l-1}{2}\rfloor}v_2$ can satisfy the edge restriction. Otherwise, there is a partition such that all $(2\lfloor\frac{l-1}{2}\rfloor+1)$ edges on the path $v_1z_1z_2\cdots z_{2\lfloor\frac{l-1}{2}\rfloor}v_2$ satisfy the edge restriction. Therefore, the effect of the path $v_1z_1z_2\cdots z_{2\lfloor\frac{l-1}{2}\rfloor}v_2$ is the same as a single edge $(v_1,v_2)$ weighted by $w(e)$. \par
(2) $c_e=0$ \par
If $v_1$ and $v_2$ belong to the same set of the partition, then all edges on the path $v_1z_1z_2\cdots z_{2\lfloor\frac{l-1}{2}\rfloor}v_2$ can satisfy the edge restriction. Otherwise, at most $2\lfloor\frac{l-1}{2}\rfloor$ edges on the path $v_1z_1z_2\cdots z_{2\lfloor\frac{l-1}{2}\rfloor}v_2$ can satisfy the edge restriction. Therefore, the effect of the path $v_1z_1z_2\cdots z_{2\lfloor\frac{l-1}{2}\rfloor}v_2$ is the same as a single edge $(v_1,v_2)$ weighted by $w(e)$. \par
In conclusion, the instance $(G_2,w',c')$ is equivalent to that of $G$, and we can recover an optimal assignment of $G$ in time $O(n)$ if we know that of $G_2$. \par
By Theorem \ref{thm71}, we can solve the instance on $G_2$ in time $O(2^k((n+k)^{3/2}\log (n+k)+k(n+k)))$. This proves the theorem.
\end{proof}

\section{Hardness at column weight three}\label{sec:hard}

\begin{proof}[Proof of Theorem \ref{thm:weight-three}]
Membership in NP is obvious. For NP-hardness, we reduce from Exact Cover by 3-Sets (X3C) \cite[Problem SP2]{GareyJohnson1979}. Let $\mathcal U$ be a universe set of size $3q$, and let $S_1,\dots,S_m$ be three-element subsets of $\mathcal U$. Take $A$ to be the element--set incidence matrix. We will prove that $\mathcal U$ has an exact cover by sets in $S_1,\dots,S_m$ if and only if there is $x\in\Ftwo^m$ such that $Ax=\one,\wt(x)\le q.$ \par
Suppose $\mathcal U$ has an exact cover by sets in $S_1,\dots,S_m$. Let $x\in\Ftwo^m$ be the vector whose coordinates corresponding to the chosen set are 1, and 0 otherwise. Then $Ax=\one$ and $\wt(x)=q$. Conversely, suppose $Ax=\one$ and $\wt(x)\le q$. Each element is covered an odd (hence at least one) number of times, so the selected sets account
for at least $3q$ incidences. Since at most $q$ sets, each of size three, are selected, they account for at most $3q$ incidences. Thus equality holds throughout: exactly $q$ sets are selected and every element occurs exactly once. These sets form an exact cover.
\end{proof}

\section{Subdivision invariance and unbounded separation} \label{separation}

\begin{proposition}[subdivision invariance]
	\label{prop:subdivision}
	Suppose $G=(V,E)$ is a graph. If $G'$ is obtained from $G$ by replacing one nonloop edge $uv$ by a nonempty $u$--$v$ path whose internal vertices are new and have no other incident edges, then
	\[
	\rctr(G')=\rctr(G).
	\]
\end{proposition}

\begin{proof}
	By Theorem \ref{thm:equality} we only need to prove $\rcex(G')=\rcex(G)$. Pick an Eulerian basis $\mc{B}$ of $G$ such that $\rho_{\cB}=\rcex(G)$. There is a natural linear isomorphism $f:\cZ(G)\to\cZ(G')$: the coordinate of the
	replaced edge is copied to every edge of the replacement path, and all other coordinates remain unchanged. Thus every Eulerian basis of one graph corresponds to an Eulerian basis of the other. Suppose the Eulerian basis of $G'$ corresponding to $\mc{B}$ is $\mc{B}'$. Then the column vectors in $A_{\mc{B}'}$ corresponding to all edges of the replacement path are the same. Moreover, the number of appearances of each subdivided edge in the new Eulerian basis is the same as the original edge in the original Eulerian basis. If that number is at most two, none of these columns belongs to $J_{\cB'}$. If it is at least three, replacing one column by several identical columns does not change the rank of the exceptional-column submatrix. Hence $\rho_{\cB'}=\rho_{\cB}$ and $\rcex(G')\le\rcex(G)$. Applying the inverse cycle-space isomorphism to an optimal Eulerian basis of $G'$ replaces the identical path columns by one column and gives the reverse inequality. Therefore $\rcex(G')=\rcex(G)$.
\end{proof}

\begin{theorem}[unbounded separation]
\label{thm:subdivision-separation}
For $L\ge1$, let $G_L$ be obtained from $K_{3,3}$ by replacing every edge by a path of length $L$.  Then
\begin{equation}
	\rctr(G_L)=1,
	\qquad
	\ctr(G_L)=L.
	\label{eq:subdivision-separation}
\end{equation}
Consequently, $\ctr(G)$ is not bounded by any function of $\rctr(G)$.
\end{theorem}

\begin{proof}
Consider $K_{3,3}$. It is nonplanar by  Kuratowski's theorem. Contract an edge $a_1b_1$ and we get a wheel, which is
planar. By Theorem \ref{thm:equality} and the obvious inequality $\rcex(G)\le\cycexc(G)$, $\rctr(K_{3,3})\le1$. By non-planarity of $K_{3,3}$, $\rctr(K_{3,3})=1$. Proposition \ref{prop:subdivision} gives $\rctr(G_L)=1$. \par
Now consider $\ctr(G_L)$. Contract all $L$ edges of one subdivided branch merges its two branch vertices and we get a subdivision of a wheel. So $\ctr(G_L)\le L$. \par
For the reverse inequality, let $P$ contain fewer than $L$ edges, and $V(G_L)$ be the vertex set of $G_L$. Any two of
the six original branch vertices have distance at least $L$, so no connected component of $(V(G_L),P)$ contains two vertices present in the original $K_{3,3}$.  Moreover, every one of the nine subdivided branches retains at least one uncontracted edge. In $G_L/P$, the six vertices present in $K_{3,3}$ are therefore distinct and joined by
nine internally disjoint nonempty paths, the same way as the incidence pattern of $K_{3,3}$. Therefore one can contract edges in $G_L/P$ and get $K_{3,3}$. By Kuratowski's theorem, $G_L/P$ cannot be planar.
\end{proof}

\end{document}